\documentclass[prd,preprint,superscriptaddress,nofootinbib,preprintnumbers]{revtex4}
\usepackage{natbib}
\pdfoutput=1

\usepackage{hyperref}
\usepackage[english]{babel}
\usepackage{amsmath,amssymb,bm,slashed,array}
\usepackage{amsfonts}
\usepackage{graphicx}
\usepackage{ifpdf}
\usepackage[sort&compress]{natbib}
\usepackage{color}
\usepackage[normalem]{ulem}
\usepackage{multirow}
\usepackage{booktabs}
\usepackage{subfigure}
\usepackage{mathrsfs}

\graphicspath{{figs/}}

\definecolor{nicered}{rgb}{0.7,0.1,0.1}
\definecolor{nicegreen}{rgb}{0.1,0.5,0.1}
\hypersetup{colorlinks,citecolor= nicegreen,linkcolor= nicered}
\definecolor{red}{rgb}{1.0, 0, 0}

\allowdisplaybreaks

\bibpunct{[}{]}{,}{n}{}{,}

\newcommand{\diag}{\text{diag}}

\newcommand{\beq}{\begin{equation}}
\newcommand{\eeq}{\end{equation}}
\newcommand{\bea}{\begin{eqnarray}}
\newcommand{\eea}{\end{eqnarray}}

\begin{document}

\def\LjubljanaFMF{Faculty of Mathematics and Physics, University of Ljubljana,
                  Jadranska 19, 1000 Ljubljana, Slovenia }
\def\LjubljanaIJS{Jo\v zef Stefan Institute, Jamova 39, 1000 Ljubljana, Slovenia}
\def\Cincy{Department of Physics, University of Cincinnati, Cincinnati, Ohio 45221, USA}
\def\Fermilab{Theoretical Physics Department, Fermilab, P.O. Box 500, Batavia, IL 60510, USA}
\def\Weizmann{Department of Particle Physics and Astrophysics, Weizmann Institute of Science, Rehovot 7610001, Israel}
\def\CERN{CERN TH-PH Division, Meyrin, Switzerland}

\preprint{CERN-PH-TH-2015-076}
\preprint{ZU-TH 08/15}
\preprint{FERMILAB-PUB-15-137-T}

% =============================================================================
\title{Dark Matter and Gauged Flavor Symmetries}

\author{Fady Bishara}              
\email[Electronic address:]{bisharfy@mail.uc.edu}
\affiliation{\Cincy}
\affiliation{\Fermilab}

\author{Admir Greljo}
\email[Electronic address:]{admir@physik.uzh.ch} 
\affiliation{Physik-Institut, Universit\"at Z\"urich, CH-8057 Z\"urich, Switzerland}

\author{Jernej~F.~Kamenik}
\email[Electronic address:]{jernej.kamenik@cern.ch}
\affiliation{\LjubljanaIJS}
\affiliation{\LjubljanaFMF}
\affiliation{\CERN}

\author{Emmanuel Stamou}              
\email[Electronic address:]{emmanuel.stamou@weizmann.ac.il}
\affiliation{\Weizmann}

\author{Jure Zupan}              
\email[Electronic address:]{jure.zupan@cern.ch}
\affiliation{\Cincy}

\date{\today}
% =============================================================================

\begin{abstract}
We investigate the phenomenology of flavored dark matter (DM). DM stability is 
guaranteed by an accidental ${\mathcal Z}_3$ symmetry, a subgroup of the standard model (SM) 
flavor group that is not broken by the SM Yukawa interactions. We consider an 
explicit realization where the quark part of the SM flavor group is fully gauged. 
If the dominant interactions between DM and visible sector are through flavor gauge 
bosons, as we show for Dirac fermion flavored DM, then the DM mass is bounded between 
roughly $0.5$\,TeV and $5$\,TeV if the DM multiplet mass is split only radiatively. 
In general, however, no such relation exists.  We 
demonstrate this using scalar flavored DM where the main interaction with the SM is through 
the Higgs portal. For both cases we derive constraints from flavor, cosmology, 
direct and indirect DM detection, and collider searches. 
\end{abstract}

%==============================================================================

\maketitle

\tableofcontents 

\newpage

%==============================================================================
%
\section{Introduction}
\label{sec:intro}
%
%==============================================================================

The stability of dark matter (DM) is commonly assumed to be due to an exact 
discrete symmetry, ${\mathcal Z}_n$. 
This can either be imposed by hand or have a dynamical origin. 
Examples include $R$-parity in the MSSM~\cite{Jungman:1995df}, 
and flavor symmetries in the leptonic sector \cite{Hirsch:2010ru,Boucenna:2011tj,Meloni:2010sk,Lindner:2011it}. 
Here, we explore the intriguing possibility raised in Refs.~\cite{Batell:2011tc,Batell:2013zwa} 
that the stability of DM is due to the $\mathcal Z_3$ center group of the global 
${\mathcal G}_F^{\rm SM}\equiv SU(3)_Q \times SU(3)_U \times SU(3)_D $ quark flavor 
symmetry. 
While ${\mathcal G}_F^{\rm SM}$ is broken by the SM Yukawa interactions, its subgroup  
$\mathcal Z_3$ remains unbroken in the SM.  
More generally, it remains exact also in the presence of New Physics (NP), if the flavor breaking 
is of Minimally Flavor Violating (MFV) type, i.e. only due to the SM Yukawas. 
The lightest neutral state that is odd under $\mathcal Z_3$ is therefore 
stable and is a DM candidate. 
This is the idea behind MFV dark matter \cite{Batell:2011tc,Batell:2013zwa,Lopez-Honorez:2013wla}.  

Requiring MFV is sufficient, but not necessary. 
In this paper we formulate a general condition for flavored DM using
{\it flavor triality} (see Eq.~\eqref{eq:flavor-triality} below). 
For example, any spurion in the bifundamental of ${\mathcal G}_F^{\rm SM}$  
leaves the above $\mathcal Z_3$ unbroken. 
The flavor breaking can thus be quite far from MFV and still have stability of DM guaranteed 
by the flavor dynamics. 
To illustrate this point we consider the model of Ref.~\cite{Grinstein:2010ve} 
where the flavor-breaking spurions have the form $Y_{u,d}^{-1}$ and are thus canonically 
not of the MFV type. 
In this model the SM quark flavor symmetry ${\mathcal G}_F^{\rm SM}$ is fully gauged
giving rise to flavor-gauge bosons,
the Yukawas are promoted to physical scalar fields (flavons) transforming under flavor, 
and in addition there is a set of chiral fermions that cancel the anomalies in the 
flavor-gauge sector. 

Using this renormalizable model we show below that a thermal relic DM can be in a nontrivial 
representation of ${\mathcal G}_F^{\rm SM}$. 
There are two conflicting constraints on this setup. 
On the one hand, Flavor Changing Neutral Current (FCNC) constraints 
impose lower bounds on the masses of states in nontrivial flavor representations. 
On the other hand, a DM relic density consistent with observations requires 
large enough DM annihilation cross section so that some of these same 
particles need to be sufficiently light. 
Both of these requirements are satisfied for ${\mathcal O}$(TeV) DM mass. 
This is low enough that it may be tested by direct and indirect 
DM detection experiments and searched for at high-energy particle colliders. 

While the phenomenology of flavored DM models can be found in 
Refs.~\cite{Agrawal:2014una,Batell:2013zwa,Lee:2014rba,Bishara:2014gwa,Agrawal:2014aoa,Agrawal:2011ze,Hamze:2014wca,Kumar:2013hfa,Kamenik:2011nb,Kile:2014jea,Kile:2011mn,Calibbi:2015sfa}, 
the construction of an explicit renormalizable model with inclusion of 
flavor-gauge interactions is new. 
Within our framework, the constraints on DM are more severe compared to a generic 
Effective Field Theory (EFT) analysis~\cite{Batell:2011tc,Lopez-Honorez:2013wla}. 
In particular, the flavor constraints from new fermionic states, and the fact that 
the vacuum expectation values (vevs) of the flavon fields need to reproduce the quark masses, makes the structure 
of the theory much more rigid and predictive. 

The paper is structured as follows. In Sec.~\ref{sec:model} we derive general 
conditions for DM to be stabilized by the exact accidental flavor symmetry of the SM (flavor triality).  
An explicit realization of this possibility is introduced in Sec.~\ref{sec:model} 
in the form of a model with fully gauged $\mathcal G_F^{\rm SM}$.  
The resulting DM, flavor, and collider phenomenology is analyzed in detail in Sec.~\ref{sec:pheno}. 
We summarize our conclusions in Sec.~\ref{sec:conclusions}, while more technical details of 
some of our computations are relegated to the Appendices.

%==============================================================================
%
\section{Stability of flavored dark matter}
\label{sec:model}
%
%==============================================================================

We start by formulating the general conditions required for flavored DM to be stable 
due to flavor triality. The SM exhibits a large global flavor symmetry
$U(3)_Q \times U(3)_U \times U(3)_D \times U(3)_L \times U(3)_E\,$ in the limit of vanishing 
Yukawa interactions.
In this paper we focus on the quark sector.  
This has the global symmetry  ${\mathcal G}_F^{\rm SM}\times U(1)_Y \times U(1)_B \times U(1)_{\rm PQ}$. 
The three $U(1)$ factors are the hypercharge, baryon number ($B$), and the Peccei-Quinn symmetry, 
respectively. 
The remaining semisimple group is ${\mathcal G}_F^{\rm SM}=SU(3)_Q \times SU(3)_U \times SU(3)_D$. 
The SM quarks transform under it  as 

\begin{align}
Q_L &\sim (3,1,1)\,, &U_R &\sim (1,3,1)\,, & D_R  &\sim (1,1,3)\,.
\end{align}
The ${\mathcal G}_F^{\rm SM}$ global symmetry is broken by the SM Yukawa terms
\begin{align}
\mathcal L_Y & =  \bar Q_L \tilde H y_{u} U_R +  \bar Q_L H y_{d} D_R + \rm h.c.\,,\label{SM:Yukawa}
\end{align}
where $\tilde H = i \sigma_2 H^*$.
${\mathcal L}_Y$ is formally invariant under  
${\mathcal G}_F$ if $y_{u,d}$ are promoted to spurions that transform as 
$(3, \bar 3,1)$ and $(3, 1,\bar 3)$ \cite{D'Ambrosio:2002ex,Chivukula:1987py,Gabrielli:1994ff,Ali:1999we,Buras:2000dm,Buras:2003jf,Kagan:2009bn}.  
NP is of MFV type if $y_{u,d}$ are the only flavor-breaking spurions also 
in the NP sector.

The SM Yukawa couplings in Eq.~\eqref{SM:Yukawa} break $U(1)_{\rm PQ}$ and 
break ${\mathcal G}_F^{\rm SM}$ to its center group ${\mathcal Z}_3^{UDQ}$, 
under which all three generations of quarks transform as

$ \{U_R, D_R, Q_L\} \to e^{i 2 \pi/3} \{U_R, D_R, Q_L\}\,$.  
In the SM the ${\mathcal Z}_3^{UDQ}$ is identical to a subgroup of $U(1)_B$. 
This is no longer true in the presence of NP. 
In MFV for instance,  ${\mathcal Z}_3^{UDQ}$ remains exact, while  
$U(1)_B$ can be broken, e.g., by dimension-$9$ operators \cite{Smith:2011rp} 
(see also \cite{Batell:2013zwa}).

The ${\mathcal Z}_3^{UDQ}$ may be the underlying reason for the stability of DM. 
To make this explicit it is useful to introduce ${\mathcal Z}_3^\chi$, 
a diagonal subgroup of $\mathcal Z_3^{UDQ}\times \mathcal Z_3^c$. 
Here, $\mathcal Z_3^c$ is the center group of color $SU(3)_c$, under 
which $ \{U_R, D_R, Q_L\} \to e^{-i 2 \pi/3} \{U_R, D_R, Q_L\}\,$. 
All the SM fields are thus $\mathcal Z_3^\chi$ singlets.  In MFV NP  
$\mathcal Z_3^\chi$ is exact, so that the lightest  $\mathcal Z_3^\chi$ 
odd particle is stable and can be a DM candidate \cite{Batell:2013zwa}. 

We generalize this observation beyond MFV. 
To this end, we introduce the notion of flavor triality~\cite{Batell:2013zwa}. 
Consider a field $X$ in the ${\mathcal G}_F^{\rm SM}$ representation 
$X \sim (n^X_Q,m^X_Q)\times (n^X_U,m^X_U)\times(n^X_D,m^X_D)$, 
where $n_i^X, m_i^X$ are the Dynkin coefficients of the corresponding 
$SU(3)_i$ group. 
We call {\em flavor triality}  the quantity
\beq\label{eq:flavor-triality}
(n_X-m_X)\,{\rm mod}\,3,
\eeq
 where $n_X=n_Q^X+n_U^X+n_D^X$ and $m_X=m_Q^X+m_U^X+m_D^X$.

The basic requirements for flavored DM to be stable due to $\mathcal Z_3^\chi$ 
are the following. 
First of all, ${\mathcal G}_F^{\rm SM}$ needs to be a good symmetry in the UV. Secondly,
${\mathcal G}_F^{\rm SM}$ needs to be broken only by spurions $\langle \Phi\rangle$ 
with zero flavor triality, $(n_{\langle \Phi \rangle}-m_{\langle \Phi\rangle})\,{\rm mod}\, 3 =0$. 
This ensures that ${\mathcal Z}_3^\chi$ is unbroken. 
(The spurions $\langle \Phi\rangle$ need to be color singlets in order not to break color.)  
The lightest ${\mathcal Z}_3^\chi$ odd state is then stable. 
If it is a color singlet it is a potential DM candidate. 
This also means that DM is in a nontrivial flavor representation with nonzero 
flavor triality, $(n_{\chi}-m_{\chi})\,{\rm mod}\, 3 \ne 0$.

The above shows that models with flavored DM can deviate significantly from MFV. 
In particular, $\mathcal Z_3^\chi$ is not broken by a vev of any field that is in 
an adjoint or in a bifundamental of ${\mathcal G}_F^{\rm SM}$.
Specifically, any function $f(y_u, y_d)$ leaves $\mathcal Z_3^\chi$ unbroken. 
More generally, additional flavor-breaking sources that 
transform as $(8,1,1)$, $(1,3,\bar 3),\dots$ may be present without spoiling DM stability. 
While the flavor structure of such NP models is not of MFV type in general, 
the stability of DM is still a consequence of an unbroken flavor subgroup. 
DM is in a nontrivial representation of the flavor group, leading to distinct 
phenomenology depending on the nature of the flavor breaking and on which flavor 
multiplet $\chi$ belongs to. 

An important starting point in the above discussion was the assumption that 
${\mathcal G}_F^{\rm SM}$ is a good symmetry in the UV. 
This is most easily achieved, if ${\mathcal G}_F^{\rm SM}$ is gauged.   
We explore this possibility in the remainder of the paper.

%==============================================================================
%
\section{Gauged flavor interactions and dark matter}
\label{sec:model}
%
%==============================================================================

We gauge the full SM quark-flavor symmetry ${\mathcal G}_F^{\rm SM}$.  
The fermionic sector is extended to cancel the anomalies of the new gauge sector.
We use the model of Ref.~\cite{Grinstein:2010ve} that allows for ${\mathcal O}({\rm TeV})$ 
flavor gauge bosons (FGBs). 
The SM Yukawas, $y_{u,d}$, arise from the vevs of new scalar fields transforming as
\begin{align}
Y_u &\sim (\bar 3,3,1)\,, &Y_d &\sim (\bar 3,1,3)\,,
\end{align}
under ${\mathcal G}_F^{\rm SM}$. 
The minimal set of chiral fermions that ensures 
anomaly cancellation of the new gauged sector is
\begin{align}
\Psi_{uR} &\sim (3,1,1)\,, & \Psi_{dR} &\sim (3,1,1)\,, & \Psi_{uL} &\sim (1,3,1)\,, & \Psi_{dL} &\sim (1,1,3)\,,
\end{align}
where the index $L$ and $R$ represents their chirality.
Together with the SM fermions they, therefore, form vector-like representations of ${\mathcal G}_F^{\rm SM}$.
The SM gauge quantum numbers of $\Psi_{uR},  \Psi_{dR}, \Psi_{uL} , \Psi_{dL}$ 
are the same as for $U_R, D_R, U_R, D_R$, respectively, i.e., they are $SU(2)_L$ singlets 
but charged under $U(1)_Y$. Because the new fermions are vector-like under the SM,
e.g, $\Psi_{uR}$ transforms like $\Psi_{uL}$ under the SM, all anomalies in the SM sector
cancel.
Remarkably, with the above fermionic content also all mixed gauge anomalies cancel. 
In fact, one could also gauge two additional flavor diagonal $U(1)$'s, $U(1)_{U}$ and $U(1)_{D}$, as well as 
$U(1)_{B-L}$, a possibility that we do not pursue further, but is discussed in
Ref.~\cite{Grinstein:2010ve}.

The Yukawa and relevant mass terms in the Lagrangian are \cite{Grinstein:2010ve}
\beq
\label{eq:mass-terms}
\begin{split}
\mathcal L_{\rm mass} &\supset \lambda_u \bar Q_L \tilde H \Psi_{uR} + \lambda'_u \bar \Psi_{uL} Y_u \Psi_{uR} + M_u \bar \Psi_{uL} U_R \\
& + \lambda_d \bar Q_L H \Psi_{dR} + \lambda'_d \bar \Psi_{dL} Y_d \Psi_{dR} + M_d \bar \Psi_{dL} D_R + {\rm h.c.},
\end{split}
\eeq
where $\lambda_{u,d}^{(')}$ are flavor-universal coupling constants and $M_{u,d}$ 
flavor-universal mass parameters. 
The mass terms in Eq.~\eqref{eq:mass-terms} mix the states $\Psi_{uL,uR}$ and $U_{L,R}$ 
forming mass eigenstates $u_i$ and $u'_i$, where $i=1,2,3$ is the generation index 
(and similarly for down-quark states). 
After electroweak symmetry breaking the masses for the two mass-eigenstate sets  are, 
in the limit $m_{u'_i}\gg m_{u_i}$, \cite{Grinstein:2010ve,Buras:2011wi}
\beq
m_{u^i}\approx \frac{v}{\sqrt2}\frac{\lambda_u M_u}{\lambda_u'\langle Y_u\rangle_i}, \qquad m_{u'_i}\approx \lambda_u' \langle Y_u\rangle_i.
\eeq
The mass matrix for the FGBs, $A_A^a$, $A=Q,U,D$ and $a=1,\dots,8$, is governed by 
the vevs of the $Y_{u,d}$ scalar fields and the gauge couplings, 
\cite{Grinstein:2010ve,Buras:2011wi}
\beq\label{eq:gauge:M2}
\big({\cal M}^2_{AB}\big)_{ab}=\frac{1}{4}g_A g_B\,{\rm Tr}\,\big[\langle Y_u \rangle\{\lambda^a,\lambda^b\}\langle Y_u\big\rangle^\dagger \big](\delta_{AB}\delta _{AQ}-2\delta_{AQ}\delta_{BU}+Q\leftrightarrow U\big)+U,u\leftrightarrow D,d,
\eeq
with $\lambda^{a}$ the Gell-Mann $SU(3)$ matrices. 
The mass matrix is $24\times24$ dimensional. 
We denote the mass-ordered eigenstates by $A^m$, $m=1,\dots,24$, 
where $A^{24}$ is the lightest one. 
The lightest gauge boson is found to be along the three diagonal $\lambda_8$ directions.
This is a consequence of $\langle Y_u \rangle$ and $\langle Y_d \rangle$ being 
almost aligned and with very hierarchical eigenvalues, where the  
$\langle Y_u \rangle_{33}$ and $\langle Y_d \rangle_{33}$ are the smallest eigenvalues.

The SM Yukawas, $y_{u,d}$, are generated after $Y_{u,d}$ obtain a vev and the 
$\Psi_i$ fields are integrated out. 
To first order in $M_{u,d}/\langle Y_{u,d} \rangle$, this gives
\begin{align}
y_{u} \simeq   \frac{\lambda_{u} M_{u}}{\lambda'_u \langle Y_{u} \rangle }\,, & &y_{d} \simeq   \frac{\lambda_{d} M_d}{\lambda'_d \langle Y_{d} \rangle }\,.
\label{eq:Yukawas}
\end{align}
The SM Yukawas, $y_{u,d}$, are non-analytic functions of the spurions $\langle Y_{u,d} \rangle$, which
signals that the theory is strictly speaking not MFV. 
Analogously, the NP states, $u'_i, d'_i$ and $A^m$, have masses that are 
non-analytic in terms of the SM Yukawas. 
However, the low-energy observables, with only the SM fields on the external legs 
can be MFV-like. 
If the $M_{u,d}/\langle Y_{u,d}\rangle$ suppressed terms are kept in Eq.~\eqref{eq:Yukawas}, 
the $y_{u,d}$ become more involved functions of $\langle Y_{u,d}\rangle^{-1}$. 
These are analytic in $\langle Y_{u,d}\rangle^{-1}$ since the effects of NP states 
decouple in the $\langle Y_{u,d}\rangle \to \infty$ limit. Similarly, the NP 
contributions to the low-energy observables $C_i$ take the form 
$\delta C_i=F(\langle Y_u\rangle ^{-1}, \langle Y_d\rangle^{-1})=\tilde F(y_u,y_d)$, with $F, \tilde F$ 
analytic functions. 
One can thus expand $\delta C_i=a_1 y_u y_u^\dagger +a_2  (y_u y_u^\dagger)^2+ b_1 y_u y_u^\dagger y_d y_d^\dagger+\cdots$, where we assumed for illustration that the transition is due to the 
left-handed quark current. 
As long as there are no large flavor-conserving ratios, 
i.e., as long as $(\lambda_uM_u/\lambda_u')/(\lambda_d M_d/\lambda_d')\ll 1/|V_{cb}|$,  
the Taylor expansion can be truncated after a few terms (see Ref.~\cite{Kagan:2009bn} for a more detailed discussion).
In this limit, the low-energy effects are of the MFV type, suppressing the FCNCs 
to acceptable levels already for NP states at the electroweak scale. 
In a numerical analysis that we perform in Appendix~\ref{app:sec:mfv}, we find 
that the expansion of the effective weak Hamiltonian in terms of SM Yukawa 
couplings can indeed still be performed for FGB contributions.

Since $\langle Y_{u,d}\rangle$ are in the bifundamental representation 
of ${\mathcal G}_F^{\rm SM}$, the ${\mathcal Z}_3^{\chi}$ remains unbroken. 
As argued above the ${\mathcal Z}_3^{\chi}$ can be used to make flavored DM 
candidates stable. 
We consider two examples: i) a fermionic DM in a vector-like representation of 
${\mathcal G}_F^{\rm SM}$ that thermalizes with the visible sector through FGBs, 
and ii) a scalar flavored DM, that interacts with the visible sector by exchanging  
FGBs as well as the Higgs.

%==============================================================================
\subsection{Fermionic flavored dark matter}
%==============================================================================
The DM in the first model is a massive Dirac fermion, $\chi$, 
in a vector-like representation of $SU(3)_U$,
\beq
\chi\sim (1,3,1)\,,
\eeq
so that no gauge anomalies are induced.  Its mass term is
\beq
{\cal L}_{\rm mass}^{\chi}=m^0_\chi\bar \chi \chi.
\eeq
Since $\chi$ is charged under ${\mathcal Z}_3^\chi$, the lightest member 
of the $\chi$ triplet is stable. 
Note that we could also gauge a larger global group 
${\cal G}_F^{SM}\times SU(3)_\chi$, with $\chi$ transforming under $SU(3)_\chi$.
That we identify $SU(3)_\chi$ with $SU(3)_U$ is a model-building choice. 

\begin{figure}
  \begin{center}
    \includegraphics[width=.39\textwidth]{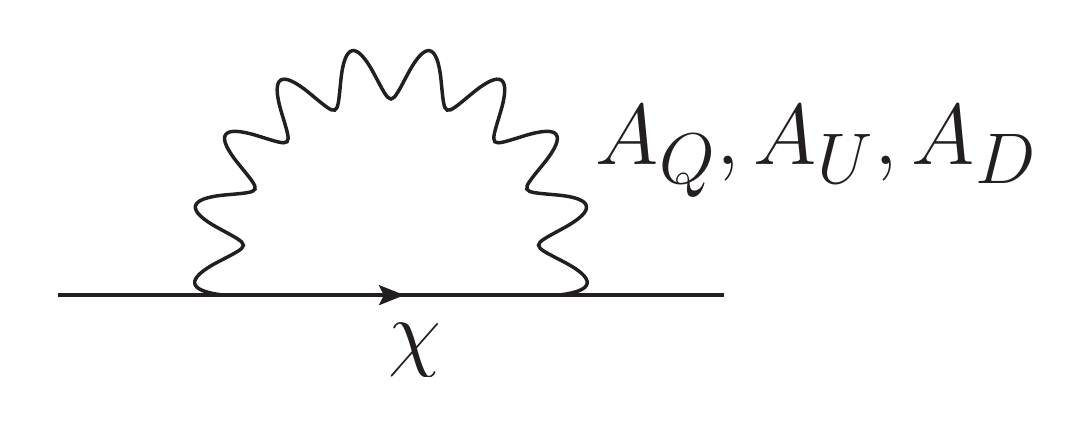}
    \\[-18mm]${}$
  \end{center}
  \caption{\label{fig:radiative_cor} Radiative corrections due to FGBs, $A_Q, A_u, A_D$,  split the DM multiplet $\chi$.}
\end{figure}

The DM flavor triplet, $\chi$, is split by radiative corrections due to the exchanges 
of FGBs, see Fig.~\ref{fig:radiative_cor}. 
In the $m^0_\chi \ll m_A$ limit, the DM mass splitting at one-loop is given by
\beq\label{DM:break}
\Delta m_\chi= -\frac{3}{4}\frac{g_U^2}{16\pi^2}m^0_\chi \Big(\Xi-\frac{1}{3}{\rm Tr}\, \Xi \Big)\,,
\eeq
where $\Delta m_\chi$ is a $3\times 3$ matrix and so is
%\beq\label{eq:xi}
$\Xi\equiv\lambda^a (\log{\cal M}_A^2/\mu^2)^{ab}\lambda^b$.
%\eeq
The FGB mass matrix ${\cal M}_A^2$ is given in 
Eq.~\eqref{eq:gauge:M2}, while the $a, b$ indices run only over the eight 
$SU(3)_U$ generators. 
The unphysical $\mu$ dependence cancels in the r.h.s.~of Eq.~\eqref{DM:break}. 
The $\chi_i$, $i=1,2, 3,$ mass eigenstates are obtained by diagonalizing the mass 
matrix $\Delta m_\chi$. 
And $\Xi$ is a function of $Y_{u}^\dagger Y_{u}$ and  $Y_{u}^\dagger  Y_{d} Y_{d}^\dagger Y_{u}$ 
vev combinations, making the $\chi$ mass eigen-basis slightly misaligned with 
respect to the up-quark one. 
The $\chi_1$ mass receives contributions from the heaviest FGBs (cf. section \ref{sec:benchmarks}). 
The lightest state is thus $\chi_1$, i.e., with the predominantly up-quark flavor, 
while the heaviest is the top-flavored state, $\chi_3$. 

\begin{figure}
  \begin{center}
    \includegraphics[width=.55\textwidth]{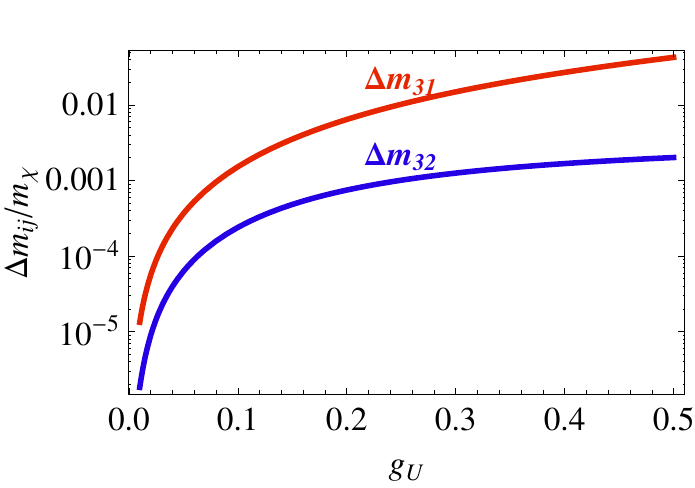}
\\[-18mm]${}$
  \end{center}
  \caption{\label{fig:mass_split_rad} 
  Typical radiative splitting of the fermionic DM multiplet with 
  $\Delta m_{31}$ ($\Delta m_{32}$) shown in red (blue) as a function of $g_U$, 
  while all other parameters are kept fixed at $g_Q=0.4, g_D=0.5$, $M_u=600$\,GeV, $M_d=400$\,GeV,
  $\lambda_u=1$, $\lambda_u'=0.5$, $\lambda_d=0.25$, $\lambda_d'=0.3$.  
  }
\end{figure}

In the numerics we use the exact one-loop expressions for the DM mass splitting, 
\beq
\Xi=\frac{3}{2}\lambda^a {\cal W}^{a+8,m} \big[B_1(m_\chi^2, m_{A_m}^2, m_\chi^2)-B_0(m_\chi^2, m_{A_m}^2, m_\chi^2)\big]{\cal W^\dagger}^{m,c+8} \lambda^c\,.
\eeq
Summation over FGB mass eigenstate indices $m=1,\dots,24$ and over $a,c=1,\dots,8$ is understood. 
The $24\times24$ dimensional matrix ${\cal W}$ diagonalizes the gauge-boson mass matrix and
$B_{0,1}$ are the Veltman-Passarino functions. 
Typical values of the mass splitting as a function of $g_U$ are shown in Fig.~\ref{fig:mass_split_rad}. 
Denoting $\Delta m_{ij}\equiv m_{\chi_i}-m_{\chi_j}$, we see that $\Delta m_{32} \ll \Delta m_{31}$, 
so that the lightest state $\chi_1$ is split away from $\chi_2$ and $\chi_3$, with the latter 
approximately degenerate, $m_{\chi_2}\simeq m_{\chi_3}$.  
This is very different from MFV DM \cite{Batell:2011tc,Batell:2013zwa,Lopez-Honorez:2013wla} 
where the DM mass splitting is assumed to be expandable in the SM Yukawa couplings. 
In that case one has an approximate $U(2)$ symmetry for the first two generations giving 
$m_{\chi_1}\simeq m_{\chi_2}$, while the top-flavored DM, $\chi_3$, is split away from 
the first two generations, and can be either significantly heavier or lighter. 

The relation $\Delta m_{32} \ll \Delta m_{31}$ signifies that our flavored DM is non-MFV. 
The flavor gauge group $SU(3)_U$ is broken by the FGB vevs $\langle Y_u\rangle$. 
This breaking is larger in the first two generations. 
Since the quark masses are inversely proportional to $\langle Y_u\rangle$, this leads to 
an approximate global $U(2)_U$ symmetry in the quark sector.  
Such an approximate symmetry is not found for the radiative corrections to DM masses, 
$m_{\chi_i}$. 
The DM multiplet has a chiral symmetry in the limit $m_\chi\to 0$ ensuring that the radiative 
corrections are proportional to $m_\chi$ and only $\log$-dependent on FGB masses. 
The splitting does not vanish in the $\langle Y_u\rangle \to \infty$ limit 
(or, equivalently, $y_u\to 0$ limit), since in this limit the $SU(3)_U$ gauge group 
is still completely broken. Numerically, for $m_\chi\sim$\,TeV the splittings are 
$\Delta m_{31}\sim {\mathcal O}(10 {\rm\,GeV})$ and $\Delta m_{32}\sim  {\mathcal O}( 1{\rm\,GeV})$ 
for $g_U\sim 0.4$ and can be less than a pion mass for an order of magnitude smaller $g_U$.

The DM multiplet can be split more significantly if there is flavor breaking 
beyond  $\langle Y_{u}\rangle$, $\langle Y_{d}\rangle$. 
As an example we consider an additional scalar field in the adjoint of 
$SU(3)_U$, $\Phi_U\sim (1,8,1)$. The DM mass Lagrangian now reads
\beq\label{DM:break:prime}
{\cal L}_{\rm mass}^{'\chi}=m^0_\chi\bar \chi_L \chi_R+\lambda_\chi\bar\chi_L \Phi_U\chi_R+ \rm h.c. \,,
\eeq
and yields DM masses that are split already at tree level, $\Delta m_\chi=\lambda_\chi \langle \Phi_U\rangle$. 
We assume that $\langle \Phi_U\rangle$ is aligned with $Y_U^\dagger Y_U$. Then the two 
diagonalize in the same basis giving ${\mathcal O}(1)$ splitting between all three members 
of the multiplet. The alignment is not needed in general, but does simplify our analysis. 
For the same reason, we also take the first state to be the lightest one, $m_{\chi_1}< m_{\chi_{2,3}}$.

The $\chi_i$ interact with the SM through FGBs. 
This also induces the decay of the heavier two states in the DM multiplet, 
$\chi_{2,3}$, to $\chi_1$.  We parametrize the relevant interactions with
\beq
\begin{split}
\mathcal{L_\chi} &\supset  (\hat g_{\chi}^m)_{ij} \bar{\chi}_i\gamma^{\mu}\chi_j A_{\mu}^{m}\\
                 & +\bar{u}_{k}\gamma^{\mu}\left( \big(\hat {\cal G}_R^u\big)_{kl,m} P_R + \big(\hat {\cal G}_L^u\big)_{kl,m} P_L \right) u_l A_{\mu}^{m}\\ 
		 & +\bar{d}_{k}\gamma^{\mu}\left( \big(\hat {\cal G}_R^d\big)_{kl,m} P_R + \big(\hat {\cal G}_L^d\big)_{kl,m}  P_L \right) d_l A_{\mu}^{m}\,,
\end{split}
\eeq
where $P_{R,L} \equiv \frac12(1\pm \gamma^{5})$ and $k,l=1,\dots,6$. 
The couplings of $\chi_i$ to the gauge bosons are 
\beq
 (\hat g_{\rm \chi}^m)_{ji} =%(U^\dagger g_{\chi}^m U)_{ji}=
 (-\frac{1}{2} g_U U^\dagger \lambda_n^U {\cal W}^{nm} U)_{ji}\,,
 \eeq
where $U$ diagonalizes the $m_\chi$ mass matrix, $U^\dagger m_\chi U=\hat m_\chi$, 
and ${\cal W}$ diagonalizes the gauge-boson mass matrix. 
The explicit form of FGB couplings to exotic and SM quarks, $\big(\hat {\cal G}_{L/R}^{u/d}\big)_{kl,m}$, 
can be found in Appendix A.2 of Ref.~\cite{Buras:2011wi}. 

The partial decay width for $\chi_{i}\to \chi_j q_l\bar q_k$ is, neglecting hadronization effects, 
\beq\label{eq:decay:width}
\Gamma(\chi_i\to\chi_j q_l \bar q_k)=
\frac{3}{(2\pi)^3}\frac{\Delta m_{ij}^5}{15}\Big[\Big|\sum_{m} (\hat g_{\rm \chi}^m)_{ij} \frac{1}{m_{A^m}^2} \big(\hat {\cal G}_L^u\big)_{kl,m}\Big|^2+L\to R\Big]\,,
\eeq
where the sum runs over the FGB mass eigenstates $m=1,\dots,24$. 
Expression \eqref{eq:decay:width} is valid in the $|\Delta m_{ij}|\ll m_{\chi_i}$ limit, 
and neglecting the quark masses.  
The above approximations are valid for all the values of parameter 
for which the correct relic abundance is obtained and the FCNC, collider and 
direct DM detection constraints are satisfied.

If the mass splitting is less than the pion mass the decay $\chi_i\to\chi_j q_l \bar q_k$ 
is kinematically not allowed. 
The heavier $\chi_i$ states then decay radiatively to $\chi_i\to \chi_j \gamma\gamma$. 
For our purposes an order of magnitude estimate of the decay width suffices. 
The naive dimensional analysis estimate gives 
\begin{equation}
\begin{split}
\Gamma\left( \chi_i\rightarrow \chi_j\gamma\gamma \right) &\sim \frac{\Delta m_{{ij}}^9}{8\pi}\frac{1}{16\pi^2}\left(\frac{\alpha}{4\pi}\right)^2 \,
	\left[
	\left|\sum_{m,f} \frac{(\hat g_\chi^m)_{ij}}{m_{A^m}^2}\frac{Q_u^2}{m_f^2}\Big(\big(\hat {\cal G}_L^u\big)_{ff,m}-\big(\hat {\cal G}_R^u\big)_{ff,m}\Big) \right|^2 
	+u\to d
	\right],
\end{split}
\label{eq:nda-rad-decay}
\end{equation}
where $Q_u=2/3$ and $Q_d=-1/3$ are the electromagnetic charges of up and down quarks. 
The sum over $m$ runs over the FGB mass eigenstates, while the sum over $f$ is over 
the SM quarks and exotic states, of mass $m_f$
(for up, down and strange quarks this needs to be replaced with $\Lambda_{\rm QCD}$).

%==============================================================================
\subsection{Scalar flavored dark matter}
\label{sec:scalar:DM}
%==============================================================================
The second model has scalar DM, $\phi_{}$, in a fundamental representation of $SU(3)_U$
\beq
\phi_{}\sim (1,3,1)\,.
\eeq
The main difference with the fermionic flavored DM from the previous subsection 
is that the scalar DM interacts with the visible sector via two different types 
of interactions. 
The first is its couplings to the FGBs, which is similar to the case of the fermionic DM. 
The second is a direct coupling to the Higgs
\beq\label{eq:Higgs:portal}
{\cal L}_{\rm int}^{\rm DM}=\lambda_H (\phi_{}^\dagger \phi_{})( H^\dagger H)\,.
\eeq 
For a thermal relic the DM annihilations proceed predominantly through the Higgs portal. 
The interactions via FGBs are subdominant except if $m_{\phi}\simeq m_A^a/2$ for some $A^a$. 
Unless this is the case, the fact that the DM carries a flavor quantum number is exhibited only through 
the multiplicity of the states.

After electroweak symmetry breaking, the DM--Higgs interactions are given by
\beq
{\cal L}_{\rm int}^{\rm DM}\supset \lambda_H \big(vh+v^2/2\big) \phi_{}^\dagger \phi_{}\,,
\eeq
and the DM mass term $m_0^2 \phi_{}^\dagger \phi_{}$ is shifted by the Higgs condensate to give the DM mass of 
\beq
m_{\phi_{}}^2=m_0^2+v^2/2\,.
\eeq

The vevs of the flavons, $\langle Y_u\rangle$ and $\langle Y_d\rangle$, split the DM multiplet at tree level 
through 
\beq\label{eq:L:kappa}
{\cal L}\supset \kappa_1 (\phi_{}^\dagger \lambda^a \phi_{}) {\rm Tr}(Y_{u}^\dagger \lambda^a Y_{u})+\kappa_2 (\phi_{}^\dagger \{\lambda^a,\lambda^b\} \phi_{}) {\rm Tr}(Y_{u}^\dagger \{\lambda^a, \lambda^b\} Y_{u})\,. 
\eeq
The spectrum is also split by radiative corrections due to FGBs. 
These are quadratically divergent and proportional to the square of the FGB mass.
In principle, it is possible to fine tune the tree-level and loop contributions 
to give almost degenerate DM flavor multiplet. 
However, given the hierarchical FGB masses, it is more likely that the DM multiplet 
is split completely, and only the lightest state is relevant for DM phenomenology. 
Depending on the signs of $\kappa_i$ in Eq.~\eqref{eq:L:kappa} the lightest $\phi$ 
component can be either top-quark or up-quark flavored. 
We choose the latter option in the numerics for easier comparison 
with the fermionic case.

%==============================================================================
%
\section{Dark matter and new physics phenomenology}
\label{sec:pheno}
%
%==============================================================================
We turn next to the phenomenology of the flavored DM models. 
We perform a scan over the parameters of the models and show that the lowest 
DM states, both for the fermionic DM, $\chi$, and the scalar DM, $\phi$, 
can be thermal relics. 
To make the scan numerically tractable we rely on several approximations 
in calculating the relic density, which we explain below. 
We also discuss the predictions for direct DM detection, and 
the constraints from FCNCs and collider searches. 

\subsection{Scan results}
In the scan we fix $\lambda_u=1$ and vary $\lambda_d \in [1/(4\pi),1]$. 
The range is chosen with the expectation that one will be able to accommodate 
both the SM top and bottom quark Yukawas as well as satisfy electroweak 
precision constraints and direct $t'$ and $b'$ searches~\cite{Grinstein:2010ve}. 
In addition we vary conservatively $\lambda'_{u,d} \in[1/(4\pi)^2,{4\pi}]$\,, 
$g_{Q,U,D} \in[1/(4\pi)^2,{4\pi}]$\,, and $M_{u,d} \in [0.2,20]$\,TeV. 
To a good approximation, the variation of $M_u$ effectively compensates the fact 
that we do not vary $\lambda_u$ as seen from Eq.~\eqref{eq:Yukawas}.
We have verified that further extending these parameter ranges does not extend 
the viable DM-model parameter space. 
For instance, the upper ranges of  $g_{Q,U,D}$ and $\lambda'_{u,d}$ already lie in 
the non-perturbative regime. 
To ensure perturbative control we require that all the FGB decay widths satisfy 
$\Gamma_{A^m}<0.5\, m_{A^m}$, and that the radiative mass splitting for the fermionic 
DM is $|\Delta m_{ij}|<0.5\, m_\chi$. 
This imposes upper bounds on $g_{Q,U,D}$ that are typically close to $\sqrt{4\pi}$.  
Similar constraints on $\lambda'_{u,d}$ are expected to follow from analogous considerations 
in the flavored Higgs sector, i.e., by requiring the total widths of the flavored scalars 
$Y_{u,d}$ to be small compared to their masses. 

\begin{figure}
  \begin{center}
    \includegraphics[width=.49\textwidth]{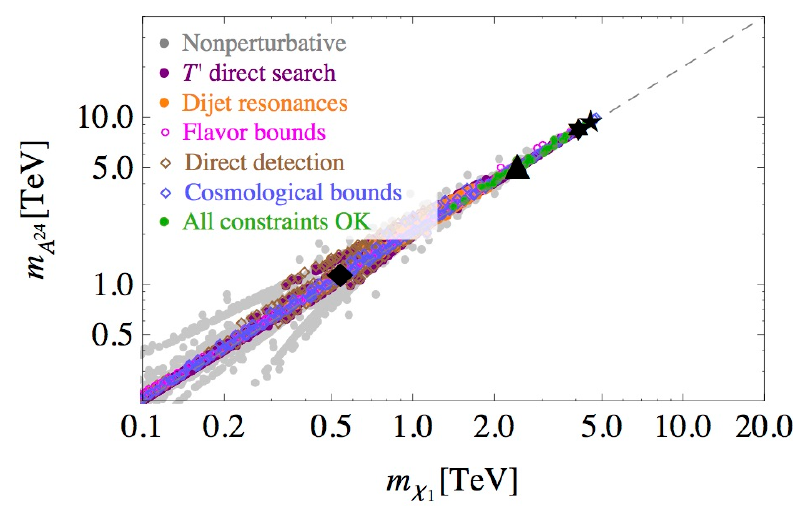}
    \includegraphics[width=.49\textwidth]{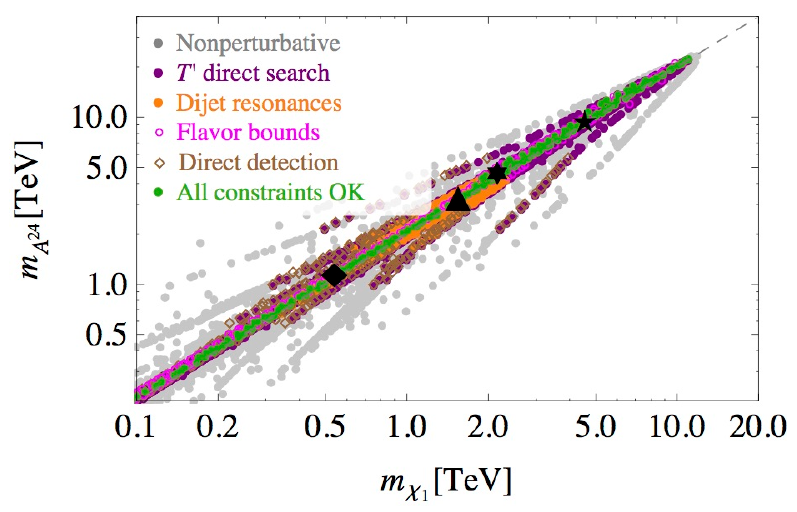}
    \includegraphics[width=.49\textwidth]{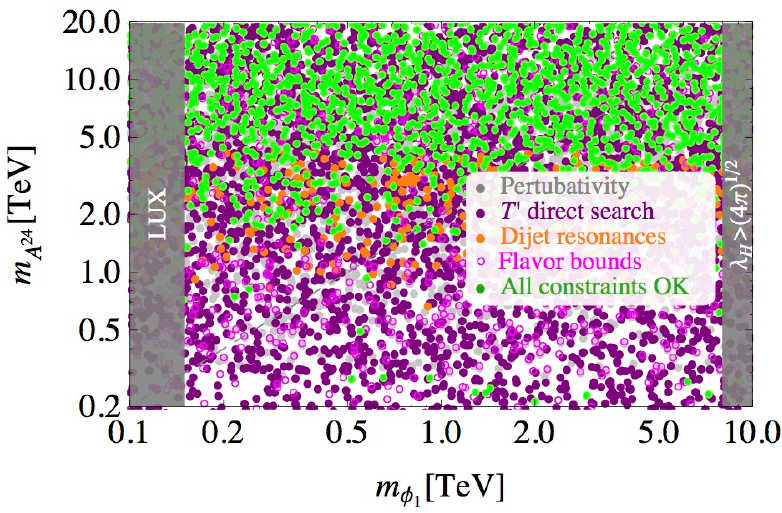}
  \end{center}
  \caption{\label{fig:mDM-fermion} The results of the scan for fermionic DM with radiative mass splitting 
  (upper left panel), in the large mass splitting limit (upper right panel) and scalar (lower panel) flavored DM.  
  Constraints from perturbativity (grey), $t'$ (dark magenta) and dijet resonance (orange) searches, 
  flavor bounds (light magenta), early-time cosmology (blue) and direct DM detection (brown) are consecutively 
  applied. Allowed parameter points are denoted by green. For scalar flavor DM (right) we show the LUX and 
  perturbativity bounds as two grey bands. The four benchmark points for fermionic flavored DM are denoted by 
  a diamond, a triangle, a hexagram and a pentagram. 
   }
\end{figure}

\begin{figure}
  \begin{center}
    \includegraphics[width=.482\textwidth]{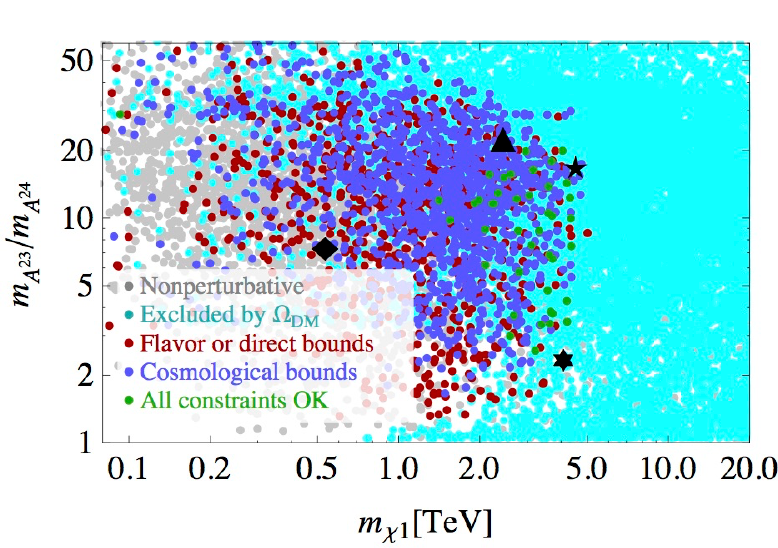}
    \includegraphics[width=.482\textwidth]{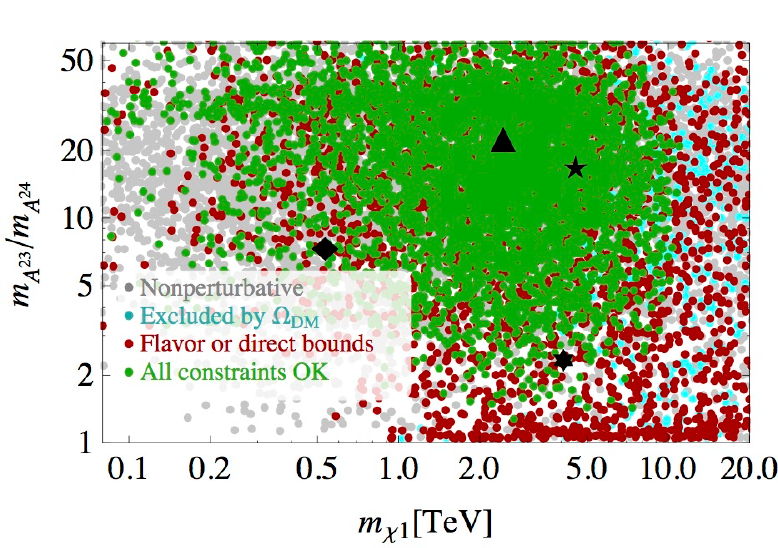}
        \includegraphics[width=.508\textwidth]{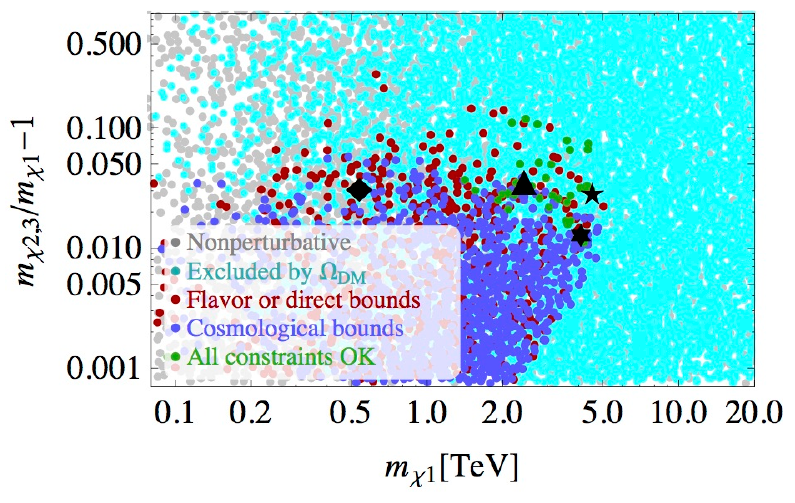}
  \end{center}
  \caption{\label{fig:mDM-fermion2} The ratio of masses of the next-to-lightest to the lightest FGBs, 
  $m_{A^{23}}/m_{A^{24}}$ for radiatively split DM multiplet (upper left panel),  and for the large mass 
  splitting limit (upper right panel), as functions of the DM mass, $m_{\chi_1}$, for the fermionic flavored DM. Lower panel shows the relative radiative mass splitting in the DM 
  multiplet.
  The constraints due to perturbativity (grey), too large relic abundance (light blue), 
  early cosmology (dark blue), flavor and direct bounds (dark red), are applied consecutively, 
  leaving allowed points (green).
  }
\end{figure}

\begin{figure}
  \begin{center}
    \includegraphics[width=.50\textwidth]{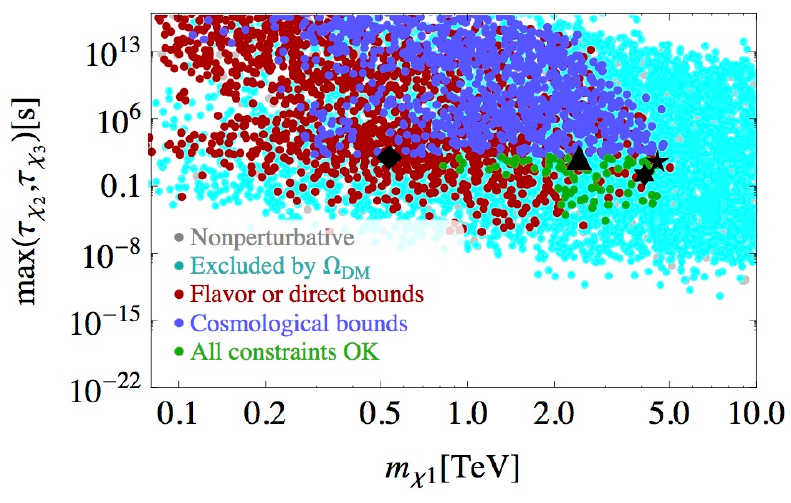}
    \includegraphics[width=.49\textwidth]{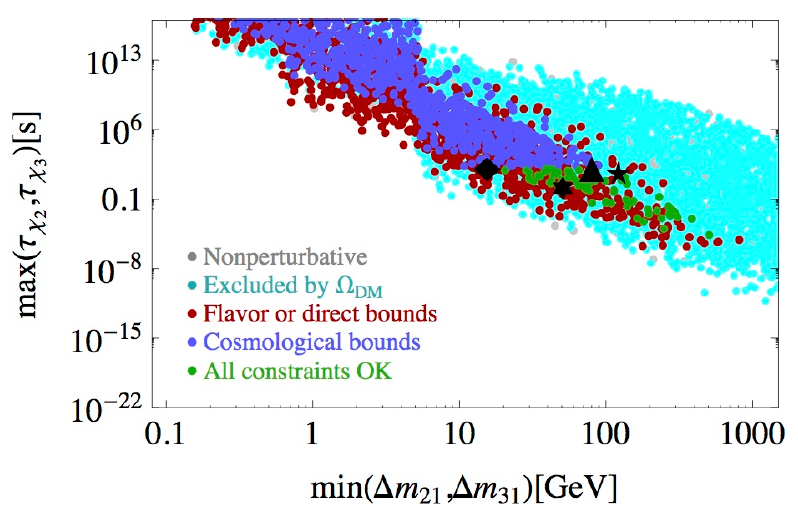}
  \end{center}
  \caption{\label{fig:mDM-fermion3} 
  The maximal decay time of the two heavy states in the DM multiplet as functions of DM mass (left) 
  and the minimal mass splitting in the DM multiplet (right) for radiatively split fermionic flavored DM. The color coding is as in Fig.~\ref{fig:mDM-fermion2}.
  }
\end{figure}

The results of the scan are shown in Figs.~\ref{fig:mDM-fermion}, \ref{fig:mDM-fermion2}, \ref{fig:mDM-fermion3}. 
Fig.~\ref{fig:mDM-fermion} (upper panels) show the results of the scan for fermionic DM model with radiative (left) 
and large tree-level mass splitting (right).  
Fig.~\ref{fig:mDM-fermion} (lower panel) instead shows the results of the scan for scalar DM. 
All the points shown in Fig.~\ref{fig:mDM-fermion} give the correct relic DM abundance, $\Omega_{\rm DM}$. 
Different colors denote consecutively applied constraints. The grey points fail the perturbativity requirement, 
$\Gamma_{A^m}<0.5 \,m_{A^m}$, $|\Delta m_{ij}|<0.5 \,m_\chi$ discussed above. 
The points in brown are excluded by direct DM detection, the points in dark magenta by $t'$ 
direct searches and the points in orange by dijet resonance searches. 
The flavor bounds exclude points in light magenta, while cosmological considerations -- mainly from 
big bang nucleosynthesis -- exclude points in dark blue. The green points are allowed by all constraints. 
In Figs.~\ref{fig:mDM-fermion2} and \ref{fig:mDM-fermion3} we also show the points where it is not possible 
to obtain the correct relic abundance (denoted by light blue), and denote by dark red the points excluded by 
the combined direct-detection, collider and flavor constraints.

\begin{figure}
  \begin{center}
    \includegraphics[width=.35\textwidth]{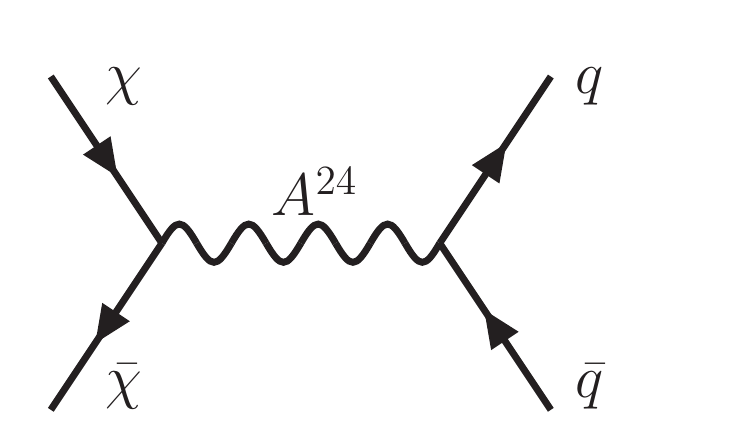}
    ~~~~~~~~
    \includegraphics[width=.35\textwidth]{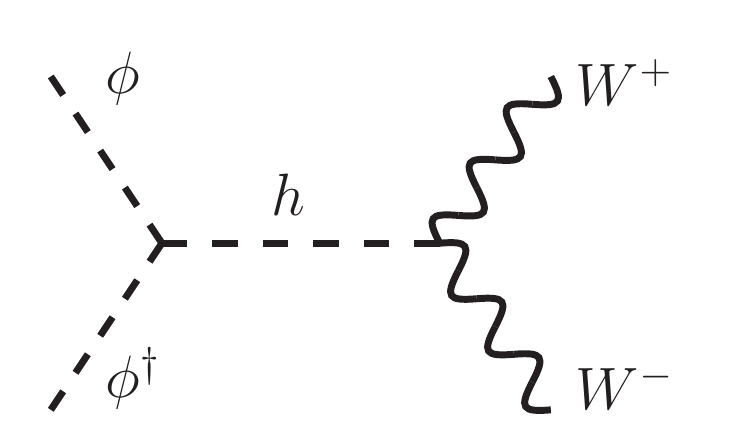}
  \end{center}
  \caption{\label{fig:DM:Feynman} 
  The Feynman diagrams for the dominant processes in the DM annihilation for 
  fermionic (left) and scalar (right) flavored DM. For scalar DM only one representative 
  diagram is shown; other relevant final states include $ b\bar b, c\bar c, \tau\tau$ and $t\bar t, hh, ZZ$ 
  (when kinematically allowed).
   }
\end{figure}

For fermionic DM the observed relic abundance requires resonantly enhanced annihilation 
through $s$-channel exchange of the lightest FGB,  $A^{24}$, see Fig.~\ref{fig:DM:Feynman}  (left). 
This leads to the correlation $m_{\chi_i}\simeq m_{A^{24}}/2$ shown in Fig.~\ref{fig:mDM-fermion} (upper panels).  
It is possible to obtain the correct relic abundance also if the DM mass is only approximately 
half of the lightest FGB (points away from the diagonal in Fig.~\ref{fig:mDM-fermion} (upper panels)). 
These points require at least some of the couplings to be large and are excluded by flavor, collider, 
direct detection, or perturbativity constraints.  
For the allowed points the relation $m_{\chi_i}\simeq m_{A^{24}}/2$ is satisfied to within a few 
decay widths of $A^{24}$, i.e. to within  ${\mathcal O}(10\%)$.
The scalar flavored DM, on the other hand, predominantly annihilates through the Higgs portal, 
see Fig.~\ref{fig:DM:Feynman}  (right). 
There is thus no relation between $m_\phi$ and $m_{A^{24}}$, as seen in 
Fig.~\ref{fig:mDM-fermion} (lower panel). 

In the remainder of this section we discuss how the various constraints 
on the DM model were obtained.

%==============================================================================
\subsection{Thermal relic}
\label{sec:thermalrelic}
%==============================================================================
For the calculation of the DM relic density we follow Refs.~\cite{Gondolo:1990dk,Griest:1990kh}. 
To speed-up the numerical scan we work in the non-relativistic limit, using the freeze-out approximation. 
The details of the calculation are given in Appendix~\ref{app:sec:thermalrelic}. 
Among viable parameter points we choose four benchmarks that satisfy all other experimental constraints. 
For the benchmark points we verify the DM relic abundance calculation using the 
{\tt MadDM}~\cite{Backovic:2013dpa} package. 
We computed the required Feynman rules using the Feynrules~\cite{Alloul:2013bka} package. 

%==============================================================================
\subsubsection{Fermionic dark matter}

In the fermionic DM model the DM annihilation to quarks is dominated by 
$s$-channel exchange of the lightest FGB, $A^{24}$, see Fig.~\ref{fig:DM:Feynman} (left panel). 
The $\chi_i\bar{\chi}_i\to\bar{u}_j u_j$ annihilation cross section  is given by
\beq\label{eq:annihilation:fDM}
\sigma(\chi_i\bar{\chi}_i\to\bar{u}_j u_j)\simeq \frac{(\hat g_{\chi}^{24})_{ii}^2 }{4\pi}\frac{s^{1/2}\left(s+2m_{\chi_i}^{2}\right)}{\sqrt{s-4m_{\chi_i}^{2}}}
\frac{\big(\hat {\cal G}_V^{u}\big)_{jj,24}^2+\big(\hat {\cal G}_A^{u}\big)_{jj,24}^2}{(s-m_{A^{24}}^{2})^{2}+m_{A^{24}}^{2}\Gamma_{A^{24}}^{2}},
\eeq
where 
\beq\label{eq:GVA:def}
\hat {\cal G}_{V,A}^{u}= \big(\hat {\cal G}_L^{u}\pm \hat {\cal G}_R^{u}\big)/2,
\eeq
$\sqrt{s}$ is the center of mass energy and  $\Gamma_{A^{24}}, m_{A^{24}}$ are the decay width 
and mass of the lightest FGB, respectively. 
In Eq.~\eqref{eq:annihilation:fDM} we have neglected quark masses; the full expression is 
given in Eq.~\eqref{app:eq:annihilation:fDM}. 
The $\chi_i\bar{\chi}_i\to\bar{d}_j d_j$ annihilation cross section follows from  
Eq.~\eqref{eq:annihilation:fDM} by replacing $u\to d$. 
The full decay width of the lightest FGB is  the sum of all partial decay widths 
for kinematically allowed channels,
\beq\label{eq:A24:decay:width}
\Gamma(A^{24}\to\bar{u}_j u_j)\simeq \frac{m_A^{24}}{4\pi}\left(\big(\hat {\cal G}_V^{u}\big)_{jj,24}^2+\big(\hat {\cal G}_A^{u}\big)_{jj,24}^2\right).
\eeq
In the above expression we have neglected the quark masses for simplicity, 
with the full expression given in Eq.~\eqref{app:eq:A24:decay:width}.
The rates for $A^{24}\to\chi_i \bar{\chi}_i,\bar{d}_j d_j$ are obtained by 
trivial coupling replacements and by correcting the color multiplicity factors. 

The correct relic abundance requires resonant annihilation, $m_\chi\simeq m_{A^{24}}/2$, 
see Fig.~\ref{fig:mDM-fermion} (upper panels). 
This implies an upper bound on the DM mass through the following argument. 
The thermally averaged DM annihilation cross section scales in the narrow width approximation as
\beq
\langle \sigma v \rangle \propto \frac{g_{A^{24}}^4}{m_{A^{24}}\Gamma_{A^{24}}}+ \mathcal O(\Gamma_{A^{24}} / m_{A^{24}})\sim \frac{1}{\langle  Y\rangle_{A^{24}}^2} \,.
\eeq
Here, we used the approximate scaling for the FGB masses and decay widths, 
$m_{A^{24}} \sim \langle  Y \rangle_{A^{24}}  g_{A^{24}}$, $\Gamma_{A^{24}} \sim (  g_{A^{24}})^2 m_{A^{24}}$.  
The $\langle Y\rangle_{A^{24}} $ and $ g_{A^{24}}$ are, respectively, the projections of 
the $Y_{u,d}$ vevs and $g_{Q,U,D}$ couplings onto the lightest FGB, $A^{24}$. 
The DM relic abundance is $\Omega_{\rm DM}\propto 1/\langle \sigma v \rangle \propto \langle Y\rangle_{A^{24}}^2$ 
and thus depends predominantly only on the flavon vevs.  
Not exceeding the relic abundance puts an upper bound 
$\langle Y \rangle_{A^{24}} \lesssim \mathcal O({\rm few~}100\,{\rm GeV})$, 
almost independent of the DM mass. Since $m_{A^{24}} \sim \langle  Y \rangle_{A^{24}}  
g_{A^{24}}$, and $ g_{A^{24}}\lesssim\sqrt{4\pi}$ for the theory to be perturbative, 
this also sets an upper bound on the lightest FGB mass. 
This in turn implies an upper bound on the DM mass through the relation 
$m_\chi\simeq m_{A^{24}}/2$. 

In the limit where only $\chi_1$ contributes to the DM relic abundance we find, 
using the scan, an upper bound $m_{\chi_1}\lesssim 10$~TeV. 
The approximation is valid if $\chi_{2,3}$ states decay well before 
$\chi_1$ freezes out (i.e. $\tau_{\chi_{2,3}} \lesssim 10^{-11}$\,s for $m_{\chi} \sim 1$\,TeV). For purely radiative DM mass splitting this is never the case (c.f. Fig.~\ref{fig:mDM-fermion3}). 
Instead, if $\chi_{2,3}$ decay after decoupling, they will also contribute to the final 
DM relic abundance and one needs to sum all three contributions. 
In this case, the constraints on the mass spectrum become much more severe. 
In particular, in order for all $\chi$ components to annihilate efficiently their masses
need to be within a few decay widths away from the lightest FGB (LFGB) mass. 
This in turn implies that the (radiative) DM mass splitting has to be of the order 
of the LFGB width. 
Since the splitting increases with $g_U$  we expect these effects to decrease 
the effective thermal DM annihilation cross section much before the theory 
becomes non-perturbative. 
Indeed we find, using the scan, an upper bound $m_{\chi_1}\lesssim 5$\,TeV.

Fig.~\ref{fig:mDM-fermion2} (upper panels) shows the ratio of masses of the next-to-lightest 
and the lightest FGB, $m_{A^{23}}/m_{A^{24}}$, as a function of DM mass $m_{\chi_1}$ for 
radiatively split DM multiplet (left) and in the large mass splitting limit (right). 
In most of the parameter space satisfying the $\Omega_{\rm DM}$ constraint 
$A^{23}$ is much heavier than $A^{24}$ so that the effects of higher resonances 
are indeed negligible. 
This justifies the use of only the lightest FGB when calculating the DM density 
in the scan. 

Fig.~\ref{fig:mDM-fermion2} (lower panel) shows the relative radiative mass 
splitting $\Delta m_{21}/m_{\chi_1}$ and $\Delta m_{31}/m_{\chi_1}$ as a function 
of the DM mass, $m_{\chi_1}$ (both splittings are shown in one plot). 
In most cases the mass splitting is below ${\mathcal O}(10\%)$ 
in order for all $\chi$ components to lie close to the resonant condition, as anticipated.  

%==============================================================================
\subsubsection{Scalar dark matter}

For scalar DM the interactions with the visible sector are mainly due to the Higgs-portal operator, 
$\lambda_H (\phi_{}^\dagger \phi_{}) (H^\dagger H)$, in Eq.~\eqref{eq:Higgs:portal}. 
The interactions due to the exchanges of FGBs are subleading except for the resonant annihilation regions $m_{\phi_1}\simeq m_{A^{24}}/2$.
By adjusting the value of $\lambda_H$ one can obtain the correct relic abundance 
for any mass of $m_{\phi_1}$ irrespective of the lightest FGB mass, $m_{A^{24}}$, 
see Fig.~\ref{fig:mDM-fermion} (bottom panel). 
In the calculation of the thermal relic abundance we include the following annihilation channels:  
$\phi_{1}^\dagger \phi_{1} \to \bar b b,\, \bar c c,\, \tau^+ \tau^-,\, W^+W^-,\, Z Z,\, h h$ and 
$\bar t t$, see Fig.~\ref{fig:DM:Feynman}. The annihilation cross sections are
\begin{align}
\sigma(\phi_{1}^\dagger \phi_{1} \to \bar f f)&=\frac{\lambda_H^2 m_f^2 N_c \left(1 - {4 m_f^2}/{s} \right)^{3/2} }{8 \pi \sqrt{1 - {4 m_{\phi_{1}}^2}/{s}} \left[(m_h^2 - s)^2 + m_h^2 \Gamma_h^2 \right]},\\
\sigma(\phi_{1}^\dagger \phi_{1} \to VV)&=\frac{c_V \lambda_H^2}{16 \pi s}\frac{ \sqrt{1 - {4 m_V^2}/{
  s}} (12 m_V^4 - 4 m_V^2 s + s^2) }{ \sqrt{
 1 - {4 m_{\phi_{1}}^2}/{s}}  \big[(m_h^2 - s)^2 + m_h^2 \Gamma_h^2\big]},
\end{align}
where $c_W=1$, $c_Z=1/2$ and
\beq
\sigma(\phi_{1}^\dagger \phi_{1} \to hh)=\frac{\lambda_H^2}{32 \pi s}\frac{ \sqrt{1 - {4 m_h^2}/{
  s}} \big[(2 m_h^2 + s)^2 + m_h^2 \Gamma_h^2\big] }{ \sqrt{
 1 - {4 m_{\phi_{1}}^2}/{s}} \big[(m_h^2 - s)^2 + m_h^2 \Gamma_h^2\big]}.
\eeq

\begin{figure}
  \begin{center}
    \includegraphics[width=.69\textwidth]{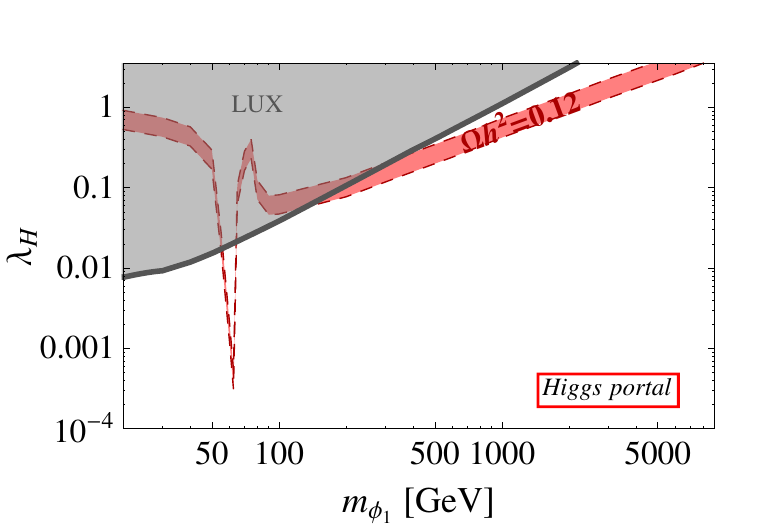}
  \end{center}
  \caption{\label{fig:higgs-portal} 
  The Higgs--DM coupling, $\lambda_H$, as a function of DM mass that gives the 
  correct relic abundance for the Higgs portal scalar DM (red band). 
  The upper (lower) dashed edge corresponds to the limit where $\phi_{2,3}$  decay much after (before) 
  the thermal freeze-out of $\phi_1$. The LUX bound, assuming correct relic abundance, 
  is shown as a shaded grey region.
  } 
\end{figure}

The thermally averaged cross sections and relic abundances are computed 
following the prescription described in Appendix~\ref{app:sec:thermalrelic}. 
The results of the scan are given in Fig.~\ref{fig:mDM-fermion} (bottom panel). 
In Fig.~\ref{fig:higgs-portal} we plot the coupling $\lambda_H$ necessary to 
obtain correct DM relic density as a function of the DM mass, $m_{\phi_{1}}$. 
As commented in Sec.~\ref{sec:scalar:DM} the flavon vevs split the lightest 
DM state $\phi_1$ from the heavier ones, such that only $\phi_1$  contributes to 
$\Omega_{\rm DM}$ (lower dashed line). 
Instead, if the splitting is too small for $\phi_{2,3}$ to decay before freeze-out, 
all three components contribute (upper dashed line).  
In both cases, requiring the Higgs-portal coupling $\lambda_H<\sqrt{4\pi}$, such
that the relic-abundance calculation is well in the perturbative regime, 
limits the DM mass $m_{\phi_{1}} \lesssim 8$\,TeV.

Note that the role of the Higgs portal may be played by other light scalars. 
In Ref.~\cite{Calibbi:2015sfa} the flavon field of the Abelian horizontal symmetry 
was used to enhance the DM annihilation cross section. 
If the flavons are light, they can also modify the phenomenology of 
the fermionic flavored DM, allowing DM annihilation into flavons.  
In this case the  DM phenomenology of the fermionic flavored DM would be 
closer to the one of our scalar DM model. 

%==============================================================================
\subsection{Cosmology}
\label{sec:cosmology}
%==============================================================================

The heavier flavored DM states, both for the fermionic DM, $\chi_{2,3}$, 
and scalar DM, $\phi_{2,3}$, are unstable. 
They decay through the $\chi_i\to \chi_j \bar q q'$ transition when the mass 
splitting is larger than the pion mass, and through the $\chi_i\to \chi_j \gamma\gamma $ 
otherwise, cf.~Eqs.~\eqref{eq:decay:width}, \eqref{eq:nda-rad-decay}. 
The SM particles in the final state of these decays can have various 
observable effects in cosmology and astrophysics.

The two relevant sets of parameters are the lifetimes of the two heavy states, 
$\tau_{\chi_{2,3}}$, and the related mass splittings of the DM multiplet (with respect 
to the lightest state), $\Delta m_{31}, \Delta m_{21}$. 
The lifetimes determine at which cosmological epoch the heavy states decay. 
The mass splittings control the released combined electromagnetic and hadronic energy, 
$E_{\rm vis} \simeq \Delta m_{21,31}$.  
They also determine the relic abundances of the heavy states. 
Generically, near the degenerate limit each state contributes roughly a third of the total DM relic abundance, $\Omega_{\rm DM}$.
Close to the resonant condition $m_{\chi} \simeq m_{A^{24}}/2$ the
$\chi_{1,2,3}$ relic abundances may differ from $\Omega_{\rm DM}/3$,
depending on the common DM mass and relative mass splittings.

For the scalar DM the mass splitting is expected to be large. 
The $\phi_{2,3}$ therefore decay before primordial nucleosynthesis. 
The decays yield negligible entropy release due to the small $\phi_{2,3}$ abundances. 
Such scenarios are basically unconstrained by current cosmological observations. 
The same is true for the fermionic DM if additional spurions split the DM multiplet 
at tree level. 

If the fermionic DM multiplet is split solely by radiative corrections, 
the $\chi_{2,3}$ and  $\chi_1$ are generically much more degenerate, cf. Fig.~\ref{fig:mDM-fermion2} (right).  
The $\chi_{2,3}$ states are then potentially long lived. 
For $\tau_{2,3}\sim {\mathcal O}(10^{-1}\,{\rm s}-10^{12}\,{\rm s})$ the decays may affect 
the primordial generation of light nuclear elements~\cite{Fields:2006ga}. 
For longer lifetimes,  $\tau_{2,3}\sim{\mathcal O}(10^{10}\,{\rm s}-10^{13}\,{\rm s})$, the $\chi_{2,3}$ 
decays distort the thermalization of the CMB by injecting high-energy photons into the 
plasma before recombination,
which is strongly constrained~\cite{Hu:1992dc,Hu:1993gc}. 
The $\chi_{2,3}$ states with lifetimes longer than $\tau_{2,3}\gtrsim 10^{10}$\,s are ruled out, 
if the injected photons carry energy above the thresholds of the efficient thermalization 
processes. 
Typically this is a fraction of the electron mass. 
For even longer lifetimes, $\tau_{2,3}\gtrsim10^{13}$\,s, the $\chi_{2,3}$ states decay after 
recombination. 
This results in photons that free-stream to us and can be searched for in 
diffuse galactic and extra-galactic gamma and X-ray spectra. 
A combination of measurements excludes scenarios with $\tau_{2,3} \lesssim 10^{26}$\,s 
all the way down to  $\Delta m_{21,31} \gtrsim \mathcal O(10)\,{\rm keV}$~\cite{Essig:2013goa}.

In the remainder of this section we consider in more detail the region 
$\tau_{\chi_{2,3}} \sim10^{-1}{\,\rm s}-10^{12}{\,\rm s}$, where the big bang 
nucleosynthesis (BBN) provides the most stringent constraints~\cite{Fields:2006ga,Iocco:2008va}. 
The injection of  energetic photons or hadrons from $\chi_{2,3}$ decays during or after BBN 
adds an additional non-thermal component to the plasma that can modify the abundances of 
the light elements~\cite{Lindley:1984bg, Reno:1987qw, Dimopoulos:1987fz, Scherrer:1987rr, Ellis:1990nb}.  
The bounds differ depending on whether the decays result in hadronic or 
electromagnetic showers in the plasma. 
The most stringent bounds are for a relic that produces mostly hadronic showers. 
This is because the electromagnetically interacting particles such as photons and electrons 
thermalize very quickly by interacting with the tail of the CMB distribution until the 
universe is $10^6$\,s old. 
In our case, the decays $\chi_{2,3}\to \chi_1 q \bar q'$ are always kinematically 
allowed for $\tau_{\chi_{2,3}}<10^{12}$\,s. 
The $\chi_{2,3}$ decays thus predominantly produce a small number of hadronic jets
with a combined released hadronic energy $E_{\rm had} \simeq E_{\rm vis}$.

There are three distinct ranges of lifetimes~\cite{Kawasaki:2004qu}. 
For $0.1\,{\rm s} \lesssim \tau_{\chi_{2,3}} \lesssim 100$\,s the dominant 
effect is the inter-conversion between protons and neutrons, which overproduces 
the ${}^4$He abundance. 
For longer lifetimes, $100\,{\rm s} \lesssim \tau_{\chi_{2,3}} \lesssim 10^7$\,s, 
hadro-dissociation is the most efficient process and the bounds come from the 
non-thermal production of Li and D. 
At late times, $10^7\,{\rm s} \lesssim \tau_{\chi_{2,3}} \lesssim 10^{12}$\,s, 
photo-dissociation caused by direct electromagnetic showers or by electromagnetic 
showers from daughter hadrons can lead to overproduction of ${}^3$He.

We impose the ${}^4$He, D and ${}^3$He constraints\footnote{The
measured ${}^4$He abundance has shifted upwards significantly since the
publicaton of Ref.~\cite{Kawasaki:2004qu}. 
This should weaken the constraints for $\tau_{2,3} \lesssim 100$\,s. 
The upward shift has no consequences for our conclusions
since we find that the ${}^4$He constraint from Ref.~\cite{Kawasaki:2004qu} 
is already never important in excluding the viable parameter space in 
our models.}
using the results in Ref.~\cite{Kawasaki:2004qu}.
The visible energy release in the decays is $E_{\rm vis} \sim \Delta m_{21,31}$.
For $100$\,GeV\,$< \Delta m_{21,31}  < 10$\,TeV  the constraints derived from the
three relic mass benchmarks in Ref.~\cite{Kawasaki:2004qu} are well approximated by a
power-law scaling with $E_{\rm vis}^{-\eta_i}$. The exponents for the three
constraints are $\eta_{{}^4\rm He} \approx 1/3$, $\eta_{\rm D} \approx 1/2$
and $\eta_{{}^3\rm He} \approx 1$. For inter-conversion and hadro-dissociation
the power-law scaling is expected to break down at energies below ${\mathcal O}(10)$\,GeV
due to the presence of hadronic thresholds~\cite{Kawasaki:2004qu}.
We thus do not extrapolate the fit results for ${}^4$He and D  for
$\Delta m_{21,31}$ below $10$\,GeV.
We assume that the photo-dissociation effects retain
approximate power law behavior for $E_{\rm vis}$ large compared to the binding
energies of the light nuclei, which is of the order of few tens of MeV.
In our model for $\tau_{\chi_{2,3}} < 10^{2(12)}$\,s,  the mass splitting,
$\Delta m_{21,31} $, is always above 10(0.1)\,GeV.
Our approximations are thus always valid for ranges of lifetimes for
which the ${}^3$He constraints are the most stringent.
For the deuterium bound, on the other hand,  the power-law scaling 
is expected to fail for part of the parameter space where the bound
is the most stringent, since $\Delta m_{21,31}$ can be as low as a few GeV.
We have checked using the power-law derived bound that these regions are
excluded by several orders of magnitude. This gives us confidence to
conclude that they remain excluded even with a more faithful treatment
of hadro-dissociation effects. 

In Fig.~\ref{fig:mDM-fermion3} we show the distribution of $\chi_{2,3}$ 
lifetimes in the viable parameter space of the fermionic DM model. 
The cosmological constraints rule out all points with $\tau_{2,3} \gtrsim 100$\,s, 
which is the range of lifetimes for which the deuterium bound becomes effective. 
The points with lifetimes  $\tau_{2,3} \lesssim 100$\,s, on the other hand, 
are never excluded by cosmological constraints. 
This is the range of lifetimes where the most strigent bound comes from the
${}^4$He abundance, which, however, is not sufficient to exclude any of our 
fermionic DM model points.

%==============================================================================
\subsection{Direct and indirect dark matter searches}
%==============================================================================
\label{sec:DM:searches}
%==============================================================================
Both fermionic and scalar flavored DM can produce direct detection signal from 
DM scattering on nuclei. 
For fermionic DM the scattering is due to $t$-channel exchanges of FGBs.  
For scalar DM the scattering is dominated by the Higgs exchange in the 
$t$-channel, while the contribution of FGBs is in general negligible.

\begin{figure}
  \begin{center}
    \includegraphics[width=.49\textwidth]{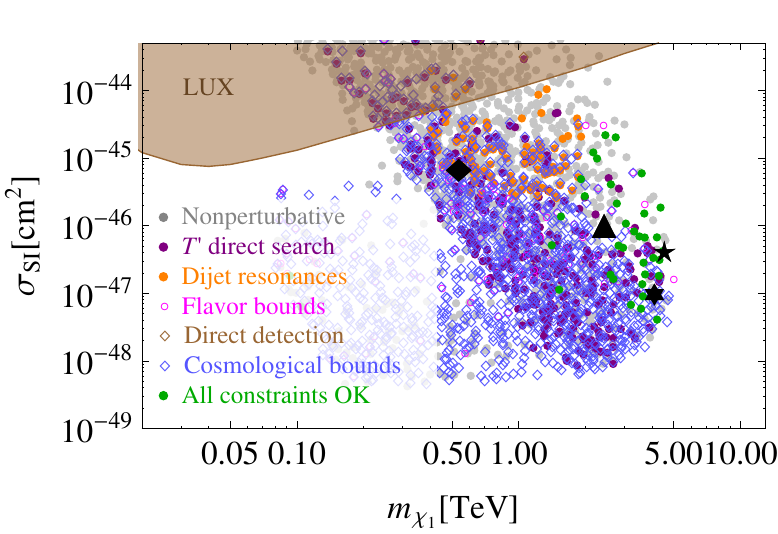}
    \includegraphics[width=.49\textwidth]{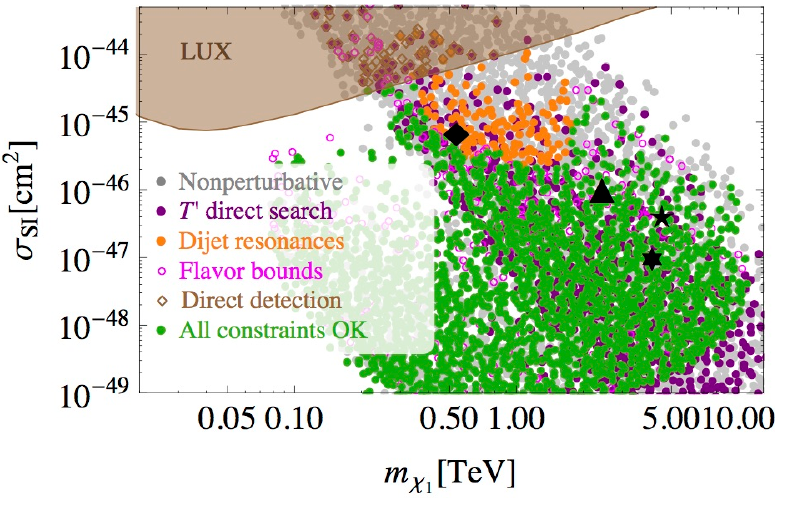}
  \end{center}
  \caption{\label{fig:directDM} 
  The predicted spin-independent cross section for DM scattering on nuclei as a 
  function of DM mass for radiatively split fermionic DM (left) and 
  in the large mass-splitting limit (right). 
  The LUX bound is the brown shaded region. 
  The color coding for the points is as in Fig.~\ref{fig:mDM-fermion}.
  }
\end{figure}

The spin-independent interactions with the nucleons for the fermionic 
flavored DM are described by the following effective 
Lagrangian~\cite{Belanger:2008sj,Arcadi:2013qia}
\beq
\mathcal{L}_{\rm dir.}= f_{p} (\bar{\chi} \gamma_{\mu}\chi)( \bar{p}\gamma^{\mu}p)+
f_{n}(\bar{\chi}\gamma_{\mu}\chi)(\bar{n}\gamma^{\mu}n)\,.
\eeq
The Wilson coefficients $f_p$ and $f_n$ are the couplings to protons and neutrons, respectively,
\beq
f_{p}=\underset{m}{\sum}(\hat g_{\chi}^m)_{11}\frac{2 \big(\hat {\cal G}_V^u\big)_{11,m} +\big(\hat {\cal G}_V^d\big)_{11,m}}{m_{A^{m}}^{2}}
\quad\textrm{and}\quad\,
f_{n}=\underset{m}{\sum}(\hat g_{\chi}^m)_{11}\frac{\big(\hat {\cal G}_V^u\big)_{11,m} +2 \big(\hat {\cal G}_V^d\big)_{11,m}}{m_{A^{m}}^{2}}.
\eeq
$\hat {\cal G}_V^{u,d}$ are the vectorial couplings of FGBs to quarks, 
defined in Eq.~\eqref{eq:GVA:def}. 
The spin-independent DM--nucleon cross section as measured by the 
LUX experiment~\cite{Akerib:2013tjd} is
\beq\label{eq:sigmaSI}
\sigma_{\chi N}^{\rm SI}=\left[1+\frac{Z}{A}\left(\frac{f_{p}}{f_{n}}-1\right)\right]^{2}\frac{\mu_{\chi n}^{2}f_{n}^{2}}{\pi},
\eeq
where $\mu_{\chi n}$ is the reduced mass of the $(\chi,n)$ system. 
The Xenon atomic and mass numbers are denoted by $Z$ and $A$, respectively.  
We thus have $Z=54$, while $A$ varies between $128$ and $134$. 
With the above relations we calculate the DM--nucleon cross section 
and compare it with the current best limits reported by the LUX 
experiment~\cite{Akerib:2013tjd}. 
The results of the scan are shown in Fig.~\ref{fig:directDM}. 
Most of the points lie well below the present LUX bound. 
This is a consequence of the fact that the relic abundance is 
given by the $s$-channel resonant annihilation, while the direct 
detection scattering is due to $t$-channel exchanges of FGBs and 
thus not resonantly enhanced.

%==============================================================================

For scalar flavored DM the dominant scattering is through $t$-channel Higgs-boson 
exchange.
This leads to the spin-independent scattering on nucleon 
$N=n,p,$~\cite{Urbano:2014hda,Cline:2012hg}
\beq
\sigma_{\chi N}^{\rm SI}=\frac{\lambda_H^2 f_{N,h}^2}{4 \pi} \left( \frac{m_{\phi_{1}} m_N}{m_{\phi_{1}} + m_N} \right)^2 \frac{m_N^2}{m_h^4 m_{\phi_{1}}^2}\,.
\eeq
The Higgs--nucleon coupling is
\beq
f_{N,h}=\frac29 +\frac79 \sum_q f_{q}^{(N)},
\eeq
where the sum runs over the light quarks, $q=u,d,s$.
$f_{q}^{(N)}$ are defined by the matrix elements of the 
light-quark scalar currents, $m_N f_{q}^{(N)} \equiv \langle N|m_q \bar q q|N\rangle$.
For the $s$ quark we use the lattice determination 
$f_s^{(N)}=0.043\pm0.011$ \cite{Junnarkar:2013ac}. 
The matrix elements for $u$ and $d$ quarks depend strongly on 
$\pi N$-scattering data.
A Baryon Chiral Perturbation Theory (B$\chi$PT) analysis of the 
$\pi N$-scattering data gives $\sigma_{\pi N} = 59(7)$\,MeV \cite{Alarcon:2011zs}. 
This is in agreement with a B$\chi$PT fit to world lattice $N_f=2+1$ QCD data, 
which gives $\sigma_{\pi N}=52(3)(8)$\,MeV \cite{Alvarez-Ruso:2014sma}. 
Including both $\Delta (1232)$ and finite-spacing parametrization 
in the fit shifts the central value to  $\sigma_{\pi N}=44$\,MeV. 
To be conservative we use $\sigma_{\pi N}=(50\pm15)$\,MeV that leads to 
$f_u^{(p)}=(1.8\pm0.5)\cdot 10^{-2}$, 
$f_d^{(p)}=(3.4\pm1.1)\cdot 10^{-2}$, $f_u^{(n)}=(1.6\pm0.5)\cdot 10^{-2}$, 
$f_d^{(n)}=(3.8\pm1.1)\cdot 10^{-2}$ 
by using expressions in Refs.~\cite{Crivellin:2014qxa,Crivellin:2013ipa}. 
This results in $f_{N,h}=(29.7\pm1.3)\cdot 10^{-2}$ where we averaged over 
Higgs couplings to proton and neutron (the difference is an order of 
magnitude smaller than the quoted error).
The resulting bound from the LUX experiment, assuming correct 
relic abundance, is shown in Fig.~\ref{fig:higgs-portal} and 
constrains $m_{\phi_{1}} \gtrsim 150$\,GeV. 

Finally, we discuss the constraints from indirect DM searches. 
The flavored DM annihilates to quarks so that the most constraining 
indirect DM searches are due to the photon and antiproton cosmic-ray fluxes. 
The constraints from the antiproton flux are quite dependent on the
propagation model. 
This can lead to almost an order of magnitude difference in uncertainty 
on the value of the excluded annihilation cross section \cite{Boudaud:2014qra}. 

For instance, by using the MED propagation model the antiproton-flux 
measurement by Pamela \cite{Adriani:2012paa} constrains the DM mass 
to be $m_\chi\gtrsim 20$\,GeV if the $\chi \chi^\dagger \to b\bar b$ annihilation dominates. 
Similar sensitivity is expected from annihilations to other quarks. 
The FERMI-LAT measurements of the photon flux from dwarf spheroidals bound 
$m_\chi\gtrsim 100$\,GeV for thermal DM annihilating to quarks \cite{Ackermann:2015zua}
(there are also slightly less stringent constraints from $\gamma$-ray 
emissions from the Large Magellanic Cloud \cite{Buckley:2015doa}, and from 
isotropic $\gamma$-ray background \cite{Ackermann:2015tah}).

\begin{figure}
  \begin{center}
    \includegraphics[width=.49\textwidth]{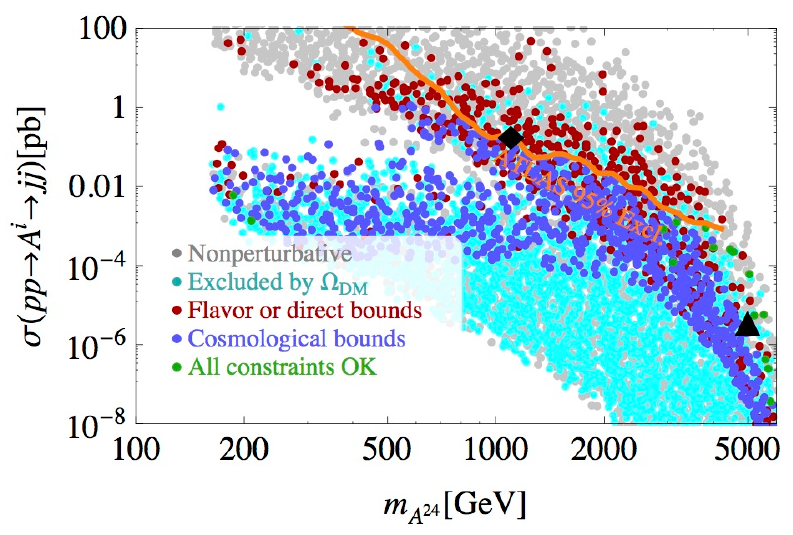}
    \includegraphics[width=.49\textwidth]{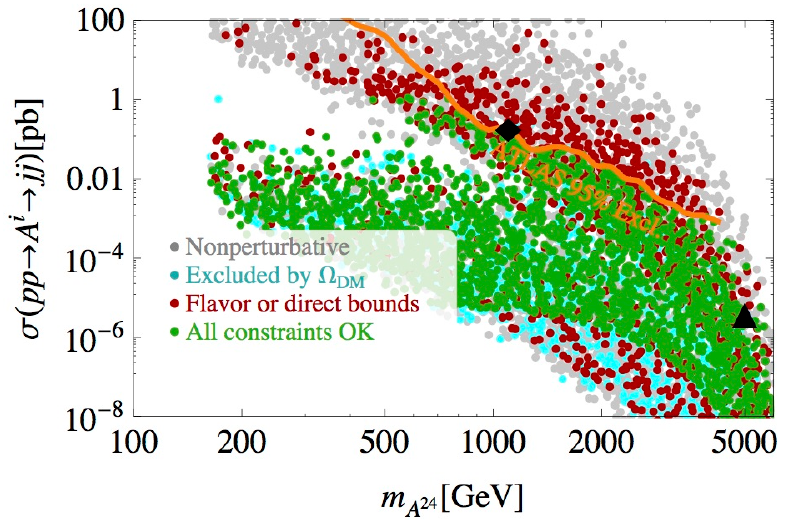}
  \end{center}
  \caption{\label{fig:dijet} The dijet production cross section at 
  $8$\,TeV LHC as a function of the lightest FGB mass for radiatively 
  split fermionic DM (left) and in the large mass splitting limit (right). 
  The 95\% CL limit from Ref.~\cite{Aad:2014aqa} is denoted with a solid 
  orange line. 
  The color coding is the same as in Fig.~\ref{fig:mDM-fermion2}.
  }
\end{figure}

%==============================================================================
\subsection{Searches at the LHC}
\label{sec:direct}
%==============================================================================
The searches for particles beyond the SM at the LHC are sensitive 
to the lightest new states in our models.
%The light states in the spectrum can be searched for directly at the LHC.
The searches for dijet resonances impose constraints on the mass of the lightest 
FGB~\cite{Aad:2014aqa}, and the searches for vector-like $T$ and $B$ 
quarks impose constraints on the mass of the lightest quark partners 
$u'_i, d'_i$~\cite{Chatrchyan:2013uxa}.

The FGBs are narrow resonances that have flavor-conserving as well as 
flavor-violating couplings to the SM quarks, $u_i,d_i$, and to the quark 
partners, $u'_i, d'_i$.
Since the FGBs are not colored they do not directly couple to gluons. 
At the partonic level the production process is dominated by 
$ q_i \bar q_j\to A^m\to q_k \bar q_l$.  
The FGBs would then appear as resonances in the dijet invariant-mass spectrum. 
For the most part, the LHC dijet resonance searches are relevant only 
for the lightest FGB which has, to a very good approximation, 
flavor-diagonal couplings to quarks. 
In this case, the cross section for $pp\rightarrow j j$ is given 
by~\cite{Eichten:1984eu,Lane:1991qh}
\newcommand{\scM}{\mathscr{M}}
\newcommand{\CL}[1]{\left|C_L^{#1}\right|^2}
\newcommand{\CR}[1]{\left|C_R^{#1}\right|^2}
\newcommand{\gvsq}[2]{\big|\hat{\cal G}_V^{#1}\big|_{#2}^2}
\newcommand{\gasq}[2]{\big|\hat{\cal G}_A^{#1}\big|_{#2}^2}
\begin{equation}
\begin{split}
\sigma(pp\rightarrow q \bar q) &=  \sum_{i,j}\int_{4 m_j^2}^{s} \frac{d\scM^2}{s}\int_{-Y_B}^{Y_B}dy_B\,\int_{-z_o}^{z_o}dz\ \,f_i\left(\sqrt{\tau}e^{y_B}\right)f_{\bar i}\left(\sqrt{\tau}e^{-y_B}\right)\frac{1}{2}
\frac{d}{dz}\hat\sigma_{ij}.
%(i\bar i\rightarrow j\bar j).
\label{eq:dijet}
\end{split}
\end{equation}
The partonic differential cross section is given by 
\begin{equation}
\begin{split}
\frac{d}{dz}\hat\sigma_{ij}
%(i\bar i\rightarrow j\bar j) 
&=
  \frac{1}{32\pi}\,\beta_f
  \frac{\scM^2}{\left(\scM^2-m_{A^m}^2\right)^2+m_{A^m}^2\Gamma_{A^m}^2}\,\left( \gvsq{u,d}{ii,m} + \gasq{u,d}{ii,m} \right)\\
  &\times\left[\left( \gvsq{u,d}{jj,m} + \gasq{u,d}{jj,m} \right)\left(1+\beta_f^2 z^2\right)%\\
  %&\qquad\qquad 
      + 4\left( \gvsq{u,d}{jj,m} - \gasq{u,d}{jj,m} \right)\,\frac{m_j^2}{\scM^2}%\\
  \right],
  %\Bigg\},
\label{eq:dijet-partonic}
\end{split}
\end{equation}
where, in the partonic center of mass frame, $\scM$ is the total energy, 
$\beta_f$ is the velocity of the final-state quarks, $z=\cos\theta^*$ is 
the cosine of the polar angle of the outgoing quark w.r.t.\ the direction 
of the incoming quark, and the couplings $\hat{\cal G}_V$, $\hat{\cal G}_A$ 
of FGBs to quarks were defined in Eq.~\eqref{eq:GVA:def}.
We have only included the $s$-channel contribution that dominates on the FGB resonance peak.
Terms odd in $z$ were dropped in the differential cross section since 
they vanish after integration.
The predicted dijet cross sections at the LHC with $\sqrt{s}=8$\,TeV are shown 
in Fig.~\ref{fig:dijet}, where the $95\%$ CL exclusion from Ref.~\cite{Aad:2014aqa} 
is denoted with a solid orange line. 
This mostly excludes the points where the lightest FGB has large couplings 
to the quarks. 
Such points are in fact already mostly excluded either by the 
perturbativity requirement or from flavor constraints.

The quark partners, $u_i',d'_i$, have an inverted mass hierarchy w.r.t.\ the SM
quarks so that in most of our scan points the $t'$ is the lightest state.
The bound on the $t^\prime$ mass depends on the $t^\prime\to bW$, $tZ$, and $th$ branching ratios.
The respective partial decay widths are given by 
\begin{align}
\Gamma(t^\prime\rightarrow bW)&=\frac{g_w^2}{64\pi}\left|s_{u_L3}V_{33}c_{d_L3}\right|^2 \frac{m_{t^\prime}^3}{m_W^2}\,\left(1-x_W^2\right)^2\left(1+2x_W^2\right),\label{eq:brtp}\\
%% BR(tprime-->tZ)
\begin{split}
\Gamma(t^\prime\rightarrow tZ)&=\frac{g_w^2}{128\pi}
  \left(c_{u_L3}s_{u_L3}\right)^2 \frac{m_{t^\prime}^3}{m_W^2}\,
  \sqrt{\left[ 1-\left( x_Z+x_t \right)^2 \right]\left[ 1-\left( x_Z-x_t \right)^2 \right]}\\
  &\hspace{12em}\times\left[\left(1-x_Z^2\right)\left(1+2x_Z^2-x_t^2\right) - x_t^2\left( 1-x_t^2 \right)\right],
  \end{split}
  \\
%% BR(tprime-->th)
\begin{split}
\Gamma(t^\prime\rightarrow th)&=\frac{\lambda_u^2}{64\pi}
  m_{t^\prime}\,
  \sqrt{\left[ 1-\left( x_h+x_t \right)^2 \right]\left[ 1-\left( x_h-x_t \right)^2 \right]}\\%\&\hspace{12em}\times
  &\qquad\times\left[
    \left( s_{u_R3}^2s_{u_L3}^2 + c_{u_R3}^2c_{u_L3}^2 \right)\left(1+x_t^2-x_h^2\right) - 
    4 s_{u_R3}\,s_{u_L3}\,c_{u_R3}\,c_{u_L3}\,x_t
  \right]\,,
  \end{split}
\end{align}
where $x_i=m_i/m_{t^\prime}$ and $s_i,~c_i$ are the sines and cosines of the 
mixing angles between the SM and exotic quarks, while $V$ is a unitary matrix 
describing the misalignment of the $Y_u$ and $Y_d$ vevs. 
Their definitions can be found in Ref.~\cite{Buras:2011wi}, where also the 
relevant Feynman rules are given. 
(We present the relevant Higgs Feynman rules in App.~\ref{app:feynrules}, correcting 
an obvious typographical error of Ref.~\cite{Buras:2011wi}.)
In Eq.~\eqref{eq:brtp} we took the limit $x_b\rightarrow 0$ that is 
justified since $m_{t'}\gg m_b$. 
We use the above expressions for the $t'\to bW, tZ, th$ branching ratios 
to obtain the 95\% confidence-level bound on the $t'$ mass by
interpolating between quoted observed-limits table in~\cite{Chatrchyan:2013uxa}. 

%==============================================================================
\subsection{Flavor constraints}
%==============================================================================
The model of gauged-flavor symmetries in Eq.~\eqref{eq:mass-terms} was designed
to be compatible with new TeV-scale dynamics and at the same time satisfy the tight
flavor constraints from FCNC observables. 
The FCNCs induced by the exchange of new states are thus relatively mild. 
The light flavor-violating gauge bosons mediate
$\Delta {\rm F}=2$ transitions at the tree-level, while the light exotic
quarks modify the loop-induced SM process. 
These modifications are large enough that they restrict the parameter
space of the model \cite{Buras:2011wi}.
All the flavor-violating parameters in the model are fixed by requiring 
$\langle Y_u \rangle $ and $\langle Y_d \rangle$ to reproduce 
the observed masses and mixings in the quark sector. 
The size of the induced FCNCs thus depend only on a relatively small set
of flavor conserving parameters in the model, the flavor symmetric masses 
and couplings.
Following the analysis in Ref.~\cite{Buras:2011wi} we focus on $\Delta
{\rm F} =2$ observables in the neutral $B$ and $K$ sectors, and  on
$\overline{B}_s\rightarrow X_s \gamma$. 

In our analysis we include the mass differences in the neutral $K^0$,
$B^0_s$, and $B^0_d$ sectors, $\Delta m_K$, $\Delta m_{B_d}$, and $\Delta m_{B_s}$,
respectively. We also include the indirect CP violation in the kaon sector, 
$\varepsilon_K$, and the mixing-induced CP asymmetries $S_{\psi K_s}$ and $S_{\psi\phi}$. 
The corresponding mixing amplitude
\begin{equation}
2\,m_M\,(M_{12}^M)^* = \langle \overline{M} \vert {\cal H}^{\Delta M=2}_{\rm eff}\vert M\rangle\,,
\label{eq:M12Mdefinition}
\end{equation}
where $M=K^0,\, B^0_d,\, B^0_s$, controls all of these observables. 

Two NP contributions to $M_{12}^M$ dominate. These are the tree-level exchanges of FGBs
and the loop-induced SM-like contribution with internal up-type quarks, including
exotic quarks. 
For the later contribution we first integrate out at the EW scale, 
$\mu_W\simeq m_W$, the exotic quarks together with the $W$ and the top 
quark. In this step we ignore the hierarchy of masses between the exotic quarks and top. 
This is a good approximation for the dominant contribution that comes from $t'$. 
The theory matches onto the EFT with the SM effective weak operators. 
We perform the Renormalization Group (RG) of the Wilson coefficients 
to the low scale at which the hadronic matrix elements are evaluated on the lattice.
For the tree-level FGB exchanges  the hard scale is given by the corresponding gauge-boson masses. 
We integrate out the FGB at the corresponding hard scale and RG evolve the Wilson coefficients
to the hadronic scale.
The FGB exchanges generate four-fermion operators with the Dirac 
structures that differ from the SM one, namely
$(\bar f_i\gamma_\mu P_R f_j)(\bar f_j\gamma^\mu P_R f_i)$ and 
$(\bar f_i\gamma_\mu P_R f_j)(\bar f_j\gamma^\mu P_L f_i)$, where $i,j$ are the flavor
indices. 
The RG evolution is implemented following Ref.~\cite{Buras:2001ra} (for further
details and the dependence of the numerical relevance with the scale see also
Ref.~\cite{Buras:2011wi}).
For the non-perturbative inputs, the decay constants and the bag parameters, we use
the current lattice averages \cite{Laiho:2009eu}.

The mass difference in the neutral kaon sector, $\Delta m_K$, and the CP-violating 
parameter $\varepsilon_K$ are given by
\begin{equation}
\Delta M_K = 2\,{\rm Re}\,(M_{12}^{K^0}),
\qquad
\varepsilon_K = \frac{\kappa_\varepsilon e^{i\varphi_\varepsilon}}{\sqrt{2}\Delta
M^{\rm exp}_K}\,{\rm Im}\,(M_{12}^{K^0})\,,
\label{}
\end{equation}
with $\varphi_\varepsilon=(43.51\pm 0.05)^\circ$ and $\kappa_\varepsilon=0.923\pm
0.006$, which includes long-distance effects in both ${\rm Im}\,M_{12}^{K^0}$ \cite{Buras:2010pza}
and in the decay, i.e.\,${\rm Im}\,\Gamma_{12}^{K^0}$ \cite{Buras:2008nn}. 
Our SM expectation for $\varepsilon_K$ incorporates the known Next-to-Next-to-Leading-Order (NNLO)
QCD corrections due to the charm \cite{Brod:2011ty} and charm--top \cite{Brod:2010mj}
contributions.

The mass differences in the neutral $B$ sectors are given by
\begin{equation}
\Delta M_{B_q}  = 2\vert M_{12}^{B_q} \vert,\quad\text{with~}q=d,s.
\label{}
\end{equation}
The CP violation in the neutral $B$ sector is probed by the time-dependent asymmetries 
in the decays $B_d^0\to\psi K_S$ and $B^0_s\to \psi\phi$ that define the observables
\begin{equation}
S_{\psi K_s }=\sin(2\beta+2\phi_{B_d})\qquad\text{and}\qquad
S_{\psi \phi}=\sin(2\vert\beta_s\vert+2\phi_{B_s})\,,
\label{}
\end{equation}
respectively. 
In the conventional parametrization of the CKM matrix the SM phases are given by 
$V_{td}^{\rm SM}=\vert V_{td}^{\rm SM}\vert e^{-i\beta}$ and 
$V_{ts}^{\rm SM}=-\vert V_{ts}^{\rm SM}\vert e^{-i\beta_s}$. The NP
phases are defined through the relation $M_{12}^{B_q}=\vert M_{12}^{B_q}\vert ~e^{2i(\beta_q + \phi_{B_q})}$.
The tree level exchanges of FGBs induce such new phases in $\Delta F=2$ matrix elements. These are thus constrained both by the $S_{\psi K_S}$ and $S_{\psi \phi}$ asymmetries, 
and by $\varepsilon_K$ in the kaon sector.

The rate of $\overline{B}\to X_s\gamma$ is also modified by the presence of
exotic up-type quarks. 
These can only enhance the $\overline{B}\to X_s\gamma$ rate with respect to the SM
expectations \cite{Grinstein:2010ve}.
The contributions of FGBs  are loop-suppressed. Even though they may be 
enhanced by $m_{b^\prime}$ they are negligible in models with a seesaw-like mass
generation for quarks,  like the model we consider \cite{Buras:2011zb}.
The SM prediction for the rate in our analysis includes the known NNLO corrections
\cite{Misiak:2006zs,Gambino:2001ew,Misiak:2006ab}.

In our numerical scan we mark parameter space points to have passed the flavor constraints 
only if the predictions for all our observables lie within three standard deviations 
of the corresponding experimental values. Whenever theoretical uncertainties are relevant, 
we include them in quadrature with the experimental ones.

The deviations of the selected FCNC observables from the SM predictions 
for the four benchmark points are shown in 
Figs.~\ref{fig:benchmarksLowMass} to \ref{fig:benchmarksMidMass2}.

%==============================================================================
\section{Benchmarks}
\label{sec:benchmarks}
%==============================================================================

\begin{figure}
  \begin{center}
   \includegraphics[height=.40\textwidth]{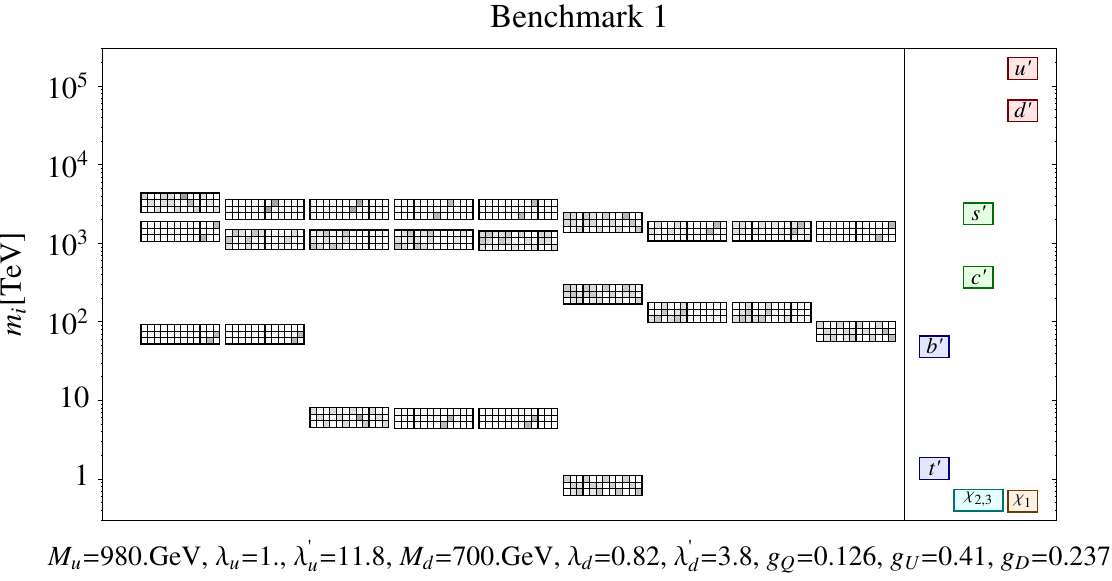}\\
   \includegraphics[width=.5\textwidth]{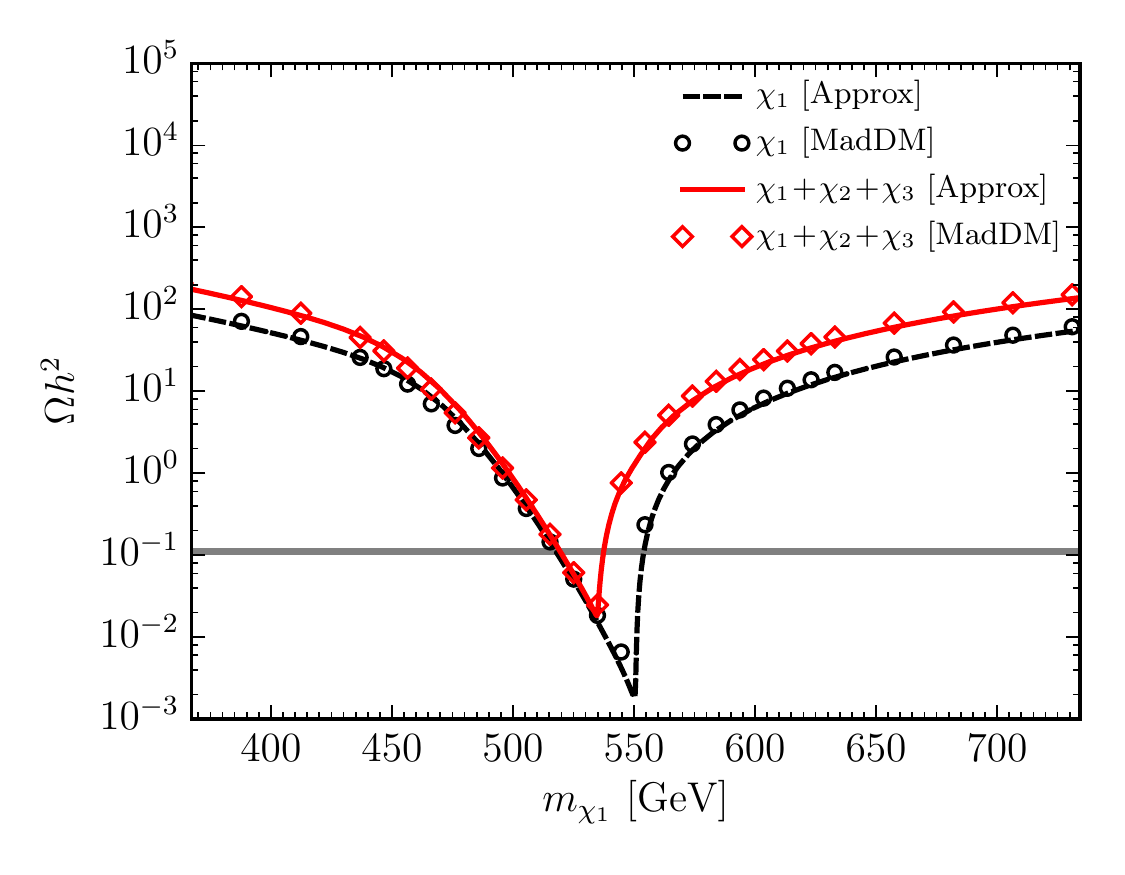}  
   \includegraphics[height=.38\textwidth]{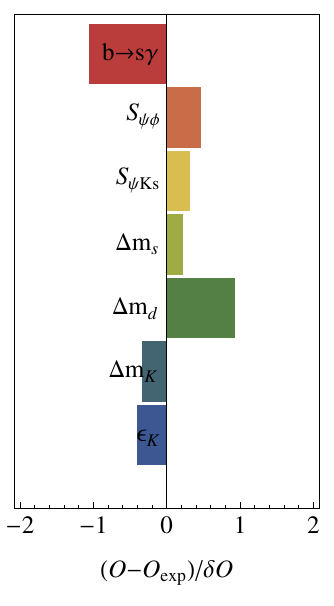}        
  \end{center}
  \caption{\label{fig:benchmarksLowMass}
  Mass spectrum and flavor decomposition (upper panel), DM relic density as a function 
  of the DM mass with all other parameters fixed (lower left panel) and the pattern of 
  effects in selected flavor observables (lower right panel) for the fermionic flavored DM benchmark 1. 
  The input benchmark-point parameters are listed in the center. See text for details.
  }
\end{figure}

\begin{figure}
  \begin{center}
    \includegraphics[height=.40\textwidth]{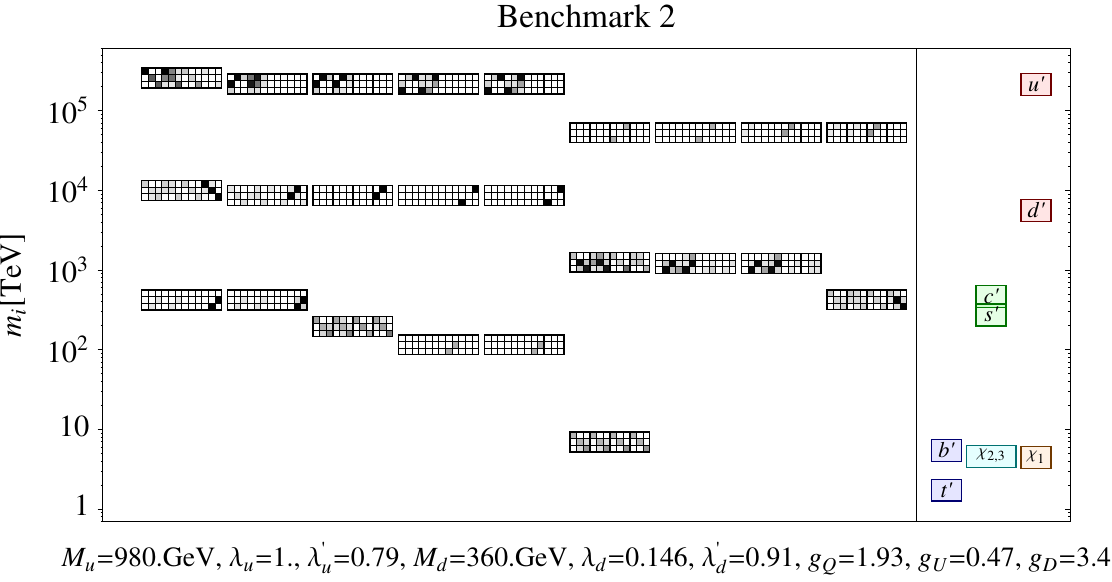}\\
  \includegraphics[width=.5\textwidth]{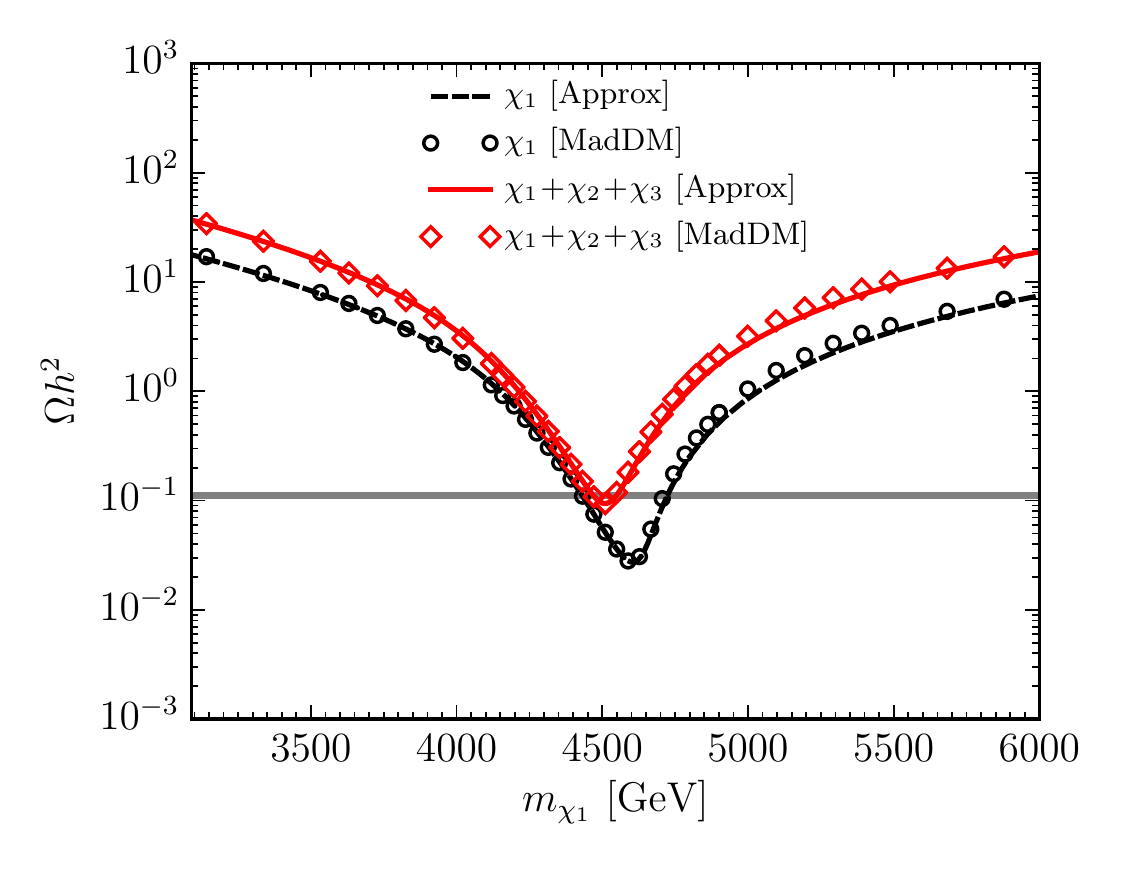}
    \includegraphics[height=.38\textwidth]{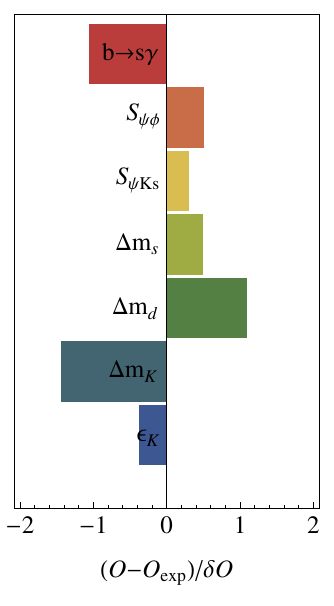}
  \end{center}
  \caption{\label{fig:benchmarksHighMass}Same as Fig.~\ref{fig:benchmarksLowMass} for benchmark 2.
 }
\end{figure}

\begin{figure}
  \begin{center}
    \includegraphics[height=.40\textwidth]{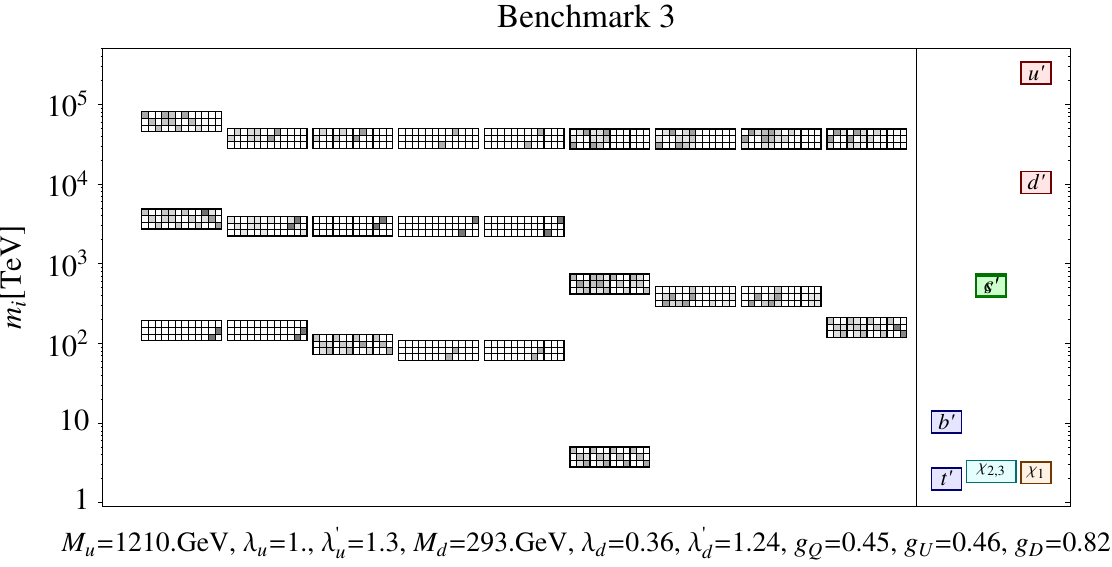}\\
    \includegraphics[width=.5\textwidth]{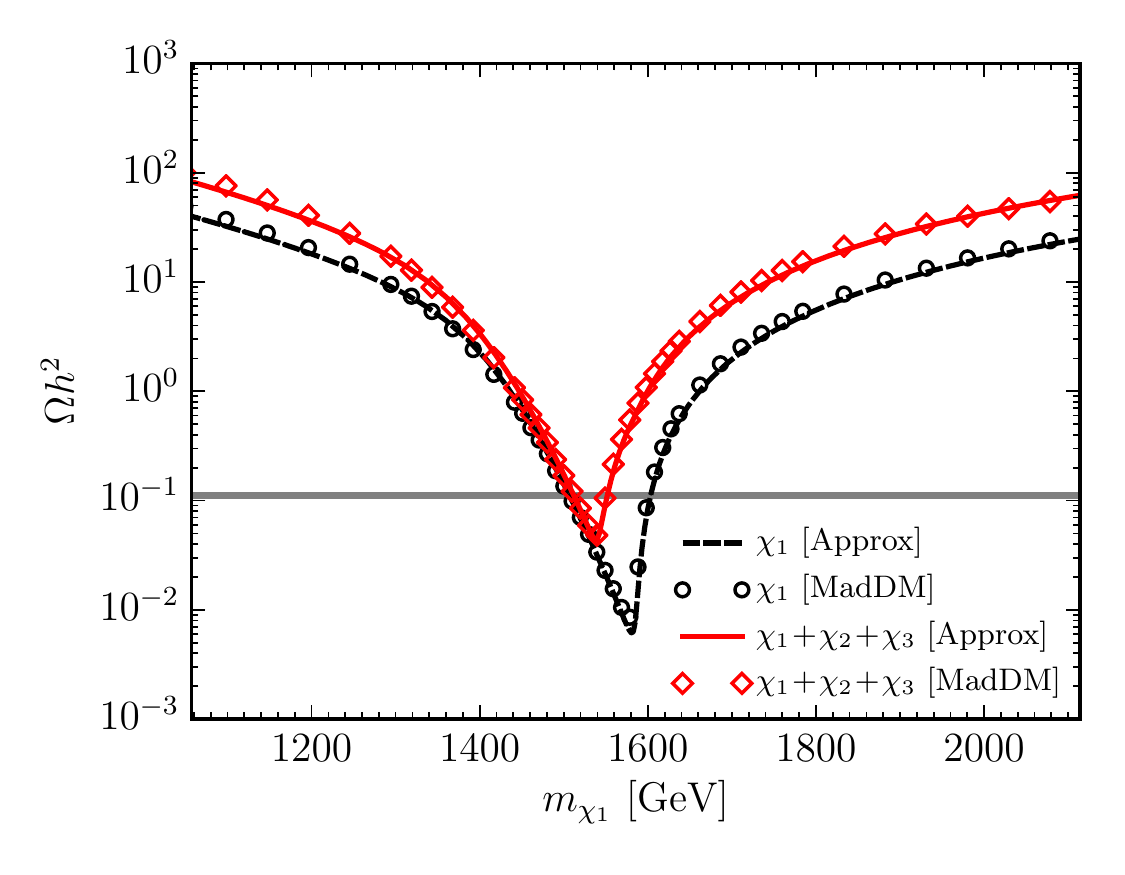}
        \includegraphics[height=.38\textwidth]{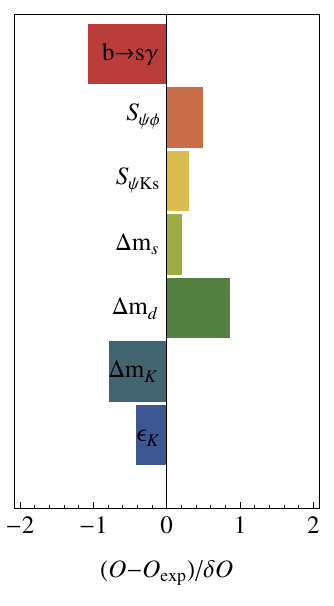}
  \end{center}
  \caption{\label{fig:benchmarksMidMass2}Same as Fig.~\ref{fig:benchmarksLowMass} for benchmark 3.}
\end{figure}

\begin{figure}
  \begin{center}
    \includegraphics[height=.40\textwidth]{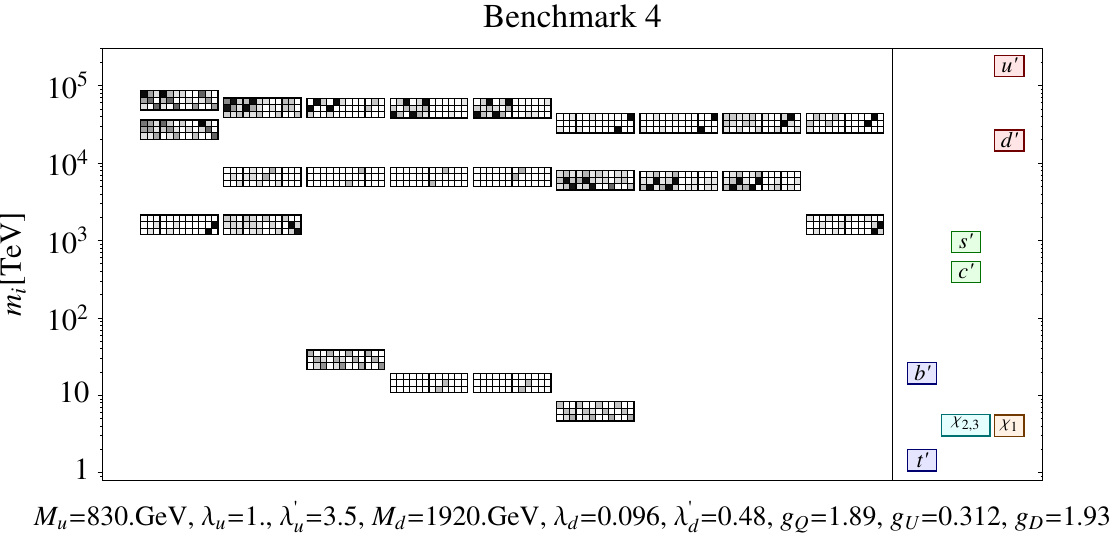}\\
  \includegraphics[width=.5\textwidth]{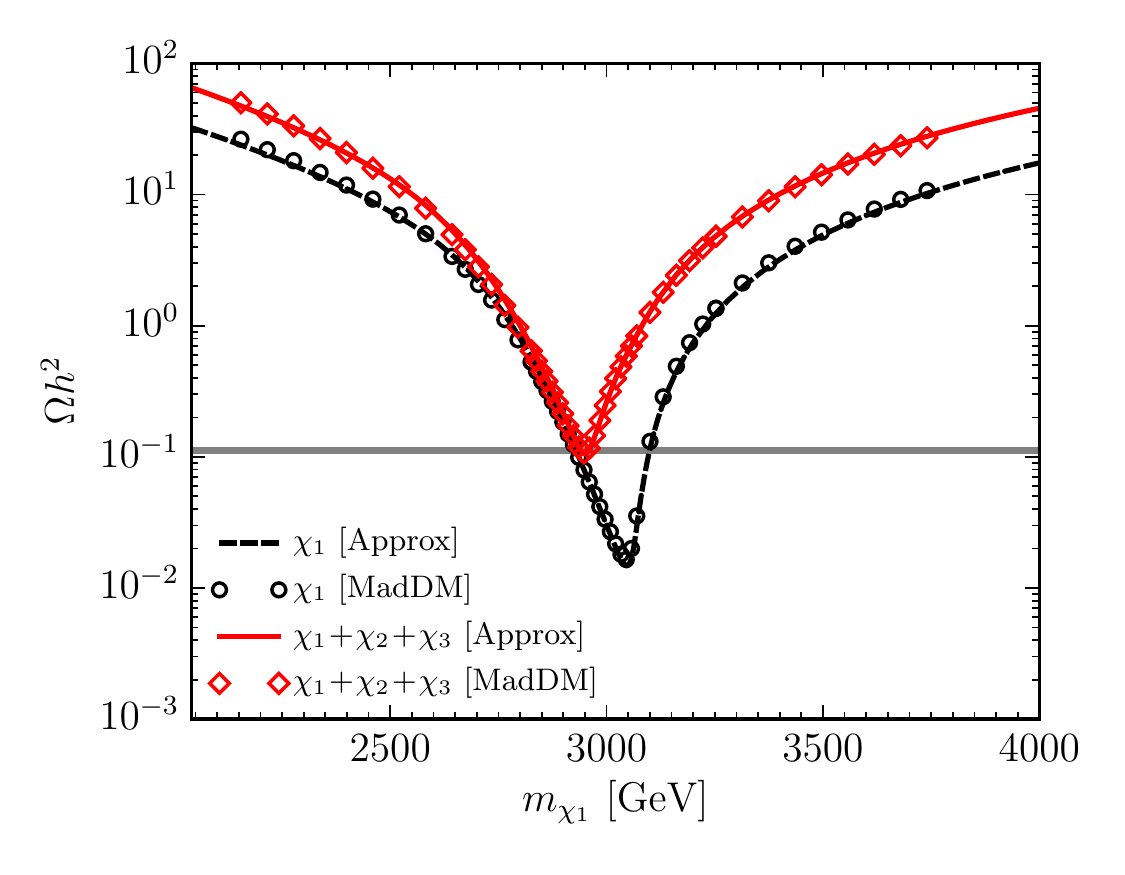}
      \includegraphics[height=.38\textwidth]{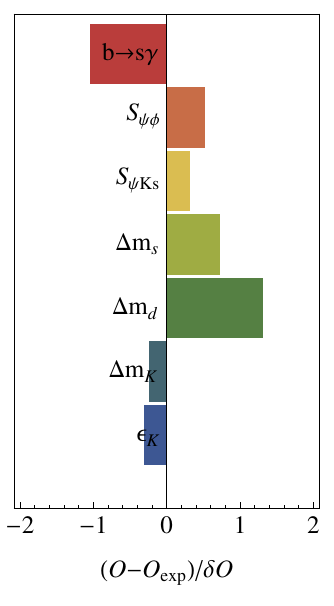}
  \end{center}
  \caption{\label{fig:benchmarksMidMass1}Same as Fig.~\ref{fig:benchmarksLowMass} for benchmark 4.
 }
\end{figure}

To illustrate the most relevant phenomenology of fermionic flavored DM we select 
four representative benchmark points. 
The main features of the four benchmarks are summarized in 
Figs.~\ref{fig:benchmarksLowMass}--\ref{fig:benchmarksMidMass1}.
The upper panels in the figures show the spectra for the FGBs, $A^m$, the quark partners, 
$u_i', d_i'$, and the DM multiplet, $\chi_i$. 
Each FGB is represented by four shaded $3 \times 3$ rasters.
The shade of the entries in the rasters is approximately logarithmically proportional 
to the size of the couplings to $u_R$, $d_R$, $u_L$ and $d_L$, respectively (from left to right). 
The DM relic abundances as functions of the $\chi_1$ mass are shown in the bottom left panels. 
The lines correspond to our approximate calculation for a radiatively 
split DM multiplet (red solid line) and for a DM multiplet with large mass splittings (black dashed line).
The open diamonds (circles) correspond to the solutions of the coupled Boltzmann equations for the radiative (large) splitting cases which were calculated in {\tt MadDM}.
The approximate and {\tt MadDM} relic-abundance calculations are in very good agreement for this small subset of benchmarks.
In general, however, a disagreement of up $\mathcal{O}(30\%)$ could be expected due to the approximations (see App.~\ref{app:sec:thermalrelic} for a more detailed discussion).
The bottom right panels show the pull in selected flavor observables, i.e., the differences 
between theoretical predictions and measurements normalized to the $1$-$\sigma$ uncertainties. 
The uncertainties were obtained by adding in quadrature the theory and experimental errors. 
The four benchmark points are also marked in Figs.~\ref{fig:mDM-fermion}, \ref{fig:mDM-fermion2}, and 
\ref{fig:mDM-fermion3} with a diamond (1), a five-point star (2), a triangle (3), and a six-point 
star (4).

{\em ``Benchmark 1''} is an example of fermionic flavored DM, where the DM multiplet is light, 
with mass below $1$\,TeV. 
The mass of the lightest state in the DM multiplet is $m_{\chi_1}\simeq 520 (540)$\,GeV if it 
lies just below (above) the LFGB resonance. 
If the mass splitting between $\chi_1$, $\chi_2$ and $\chi_3$ is solely due to radiative corrections, 
$\chi_{2}$ and $\chi_{3}$ are almost mass degenerate with masses roughly $10$\,GeV above $m_{\chi_1}$, 
and $\chi_3$ is about $100$\,MeV heavier than $\chi_2$. 
The lightest quark partner is $t'$ with a mass $m_{t'}\simeq1.3$\,TeV. 
The lightest FGB has a mass $m_{A^{24}}\simeq1.1$\,TeV. 
All the remaining NP states are above $7$\,TeV. 
This benchmark point demonstrates that even parameter regions with low lying FGBs can be consistent 
with both the resonance searches and the FCNC bounds. 
The most robust constraints in this parameter region come from cosmology 
(in case of radiatively-split DM masses) and dijet-resonance searches 
(see Figs.~\ref{fig:mDM-fermion3} and~\ref{fig:dijet}). 
Note in particular that for the completely (mass) decoupled fermionic DM scenario, 
in which cosmology bounds are absent, all experimental constraints can be satisfied 
even for DM (and LFGB) masses below few $100$\,GeV.

The bottom left panel in Fig.~\ref{fig:benchmarksLowMass} shows the predicted relic 
abundance for this benchmark, if only the DM mass is varied, which also modifies
the splitting within the DM triplet.
Relic abundance consistent with observations is obtained for a mass of DM close 
to half of the mass of the lightest FGB, in which case the annihilation cross section 
is resonantly enhanced. 
To saturate the observed DM relic density, two solutions for $m_{\chi_1}$ are obtained, 
with $m_{\chi_1}$ either above or below the resonant peak. 
We see that for radiatively split DM masses, where all $\chi_i$ components contribute 
to the DM relic density, $m_{\chi_i}$ need to lie within ${\mathcal O}(3\%)$ of the 
resonant peak for the annihilation to be strong enough. 
For completely decoupled DM multiplet the resonant condition is relaxed and 
needs to be satisfied to ${\mathcal O}(5\%)$. 

In Fig.~\ref{fig:benchmarksLowMass} (upper  panel) we show the spectrum for the lower 
mass solution and radiative DM multiplet splitting. 
We see that the quark partners of the lighter generations are heavier than the partners of
the third generation quarks. 
Similarly, the FGBs that couple more strongly to the first two generations are typically heavier 
than the ones that couple preferably to the third generation. 
The couplings of the lightest FGB to the light quarks have the form
$\hat {\cal G}_L^u\simeq \hat {\cal G}_R^u\simeq\hat {\cal G}_L^d\simeq \hat {\cal G}_R^d\propto \lambda^8$, 
where the relative corrections to this relation are below the percent level.
This means that the lightest FGB couples to the light quarks vectorially, 
$\hat {\cal G}_{A}^{u,d}\ll \hat {\cal G}_{V}^{u,d}$, to a very good approximation. 
The same is true for the majority of parameter-space points passing flavor constraints.

The largest effects in flavor physics are in the mixing observables, the mass splittings 
$\Delta m_{d,s}$ in $B_d-\bar B_d$ and $B_s-\bar B_s$ systems, respectively, and the mass 
splitting in the $K-\bar K$ mixing, $\Delta m_K$, and the related CP violating parameter $\epsilon_K$. 
The pulls in $b\to s\gamma$ and $B_d-\bar B_d$ mixing are due to the fact that the measurements 
agree with the SM prediction only at $1$-$\sigma$ level and the contribution to them from new states 
is very small. 

{\it ``Benchmark 2''} is an example of a generic parameter-space region, but towards the upper 
end of the perturbatively allowed region. 
The DM has a mass $m_{\chi_1}\simeq4.5$\,TeV, while the heavier states in the DM multiplet 
have masses $120$\,GeV and $150$\,GeV above $m_{\chi_1}$ (for the case of only radiative mass splitting). 
The lightest exotic quark is the top partner with mass $m_{t'}\simeq1.7$\,TeV, while the mass 
of the lightest FGB is $m_{A^{24}}\simeq 9.2$\,TeV. 

For such high DM masses it is barely possible to obtain the correct relic abundance
(see the lower left panel in Fig.~\ref{fig:benchmarksHighMass}). 
Therefore, the DM mass is finely tuned to be exactly on the resonant peak 
(see the lower left panel in Fig.~\ref{fig:benchmarksHighMass}), so $m_{\chi_i}\simeq m_{A^{24}}/2$. 
Because of the high masses of the NP states the direct searches 
(direct DM detection, $t'$ searches and dijet resonance searches) as well as the indirect 
flavor constraints are easily avoided, although  $K-\bar K$ mixing does receive non-negligible 
contributions. 

Also in this case, the couplings of the lightest FGB to the light quarks have the form 
$\hat {\cal G}_L^u\simeq \hat {\cal G}_R^u\simeq\hat {\cal G}_L^d\simeq \hat {\cal G}_R^d\propto \lambda^8$, 
so the couplings of the lightest FGB to quarks are to a good extent vectorial.

{\it ``Benchmark 3"} is an example of a generic parameter space in which all the couplings of 
the model are well below the pertubativity bounds. The lightest FGB has a mass $m_{A^{24}}\simeq 5$\,TeV 
while all other FGBs have masses above $100$\,TeV. The lightest partner quark is $t'$ with mass 
$m_{t'}\simeq 2$\,TeV. The DM states have masses $m_{\chi_1}\simeq2.4$\,TeV and 
$m_{\chi_2}\simeq m_{\chi_3}=2.5$\,TeV (for radiative mass splitting). 
All direct experimental constraints as well indirect flavor bounds are easily satisfied in this case. 
For radiatively DM mass splitting the cosmological constraints are the most constraining. 
In particular, requiring small enough $\tau_{\chi_{2,3}}$ (or equivalently large enough $\Delta m_{21,31}$) 
typically imposes a lower bound on $g_U$\,.

{\em ``Benchmark 4''} is an example of the case in which the next-to-lightest FGBs have masses 
not too far from the lightest FGB. 
In the benchmark point the lightest FGB has a mass $m_{A^{24}}\simeq 8.3$\,TeV, while the next 
to lightest FGBs have masses $m_{A^{23}}\simeq m_{A^{22}}\simeq 19$\,TeV, $m_{A^{21}}\simeq 39$\,TeV. 
In this case the deviations from the
$\hat {\cal G}_L^u\simeq \hat {\cal G}_R^u\simeq\hat {\cal G}_L^d\simeq \hat {\cal G}_R^d\propto \lambda^8$ 
relation for the lightest FGB coupling to quarks are of ${\mathcal O}(10\%)$. 
Nonetheless, this does not have a significant effect on the computation of the DM relic abundance. This is 
demonstrated in Fig.~\ref{fig:benchmarksMidMass1} bottom left panel where a comparison  is shown 
between the relic-abundance calculation neglecting (dashed lines, labelled ``Approx") and including 
(full lines, using MadDM) flavor off-diagonal lightest FGB couplings and contributions of heavier FGBs.
For a more detailed discussion of these effects see Appendix~\ref{app:sec:thermalrelic}.

The lightest quark partner is $t'$ with mass $m_{t'}\simeq 1.4$\,TeV, and is significantly 
lighter than all FGBs and also DM. 
The DM states are degenerate to a good approximation, with masses $m_{\chi_1}=4.1$\,TeV 
and $m_{\chi_2}\simeq m_{\chi_3}\simeq4.2$\,TeV. 
Because of the light $t'$ the flavor constraints are nontrivial, and there are visible 
effects in $B_d$ and $B_s$ mixing observables. 

%==============================================================================
%
\section{Conclusions}
\label{sec:conclusions}
%
%==============================================================================
We investigated the possibility that DM is in a nontrivial representation of the 
continuous flavor group ${\cal G}_F^{\rm SM}=SU(3)_Q\times SU(3)_U\times SU(3)_D$. 
The two main results are that (i) one can have a viable model of DM where DM is stable 
because it is charged under ${\mathcal Z}_3^\chi$ -- a discrete central subgroup of ${\cal G}_F^{\rm SM}$ 
and color $SU(3)$, and (ii) that the DM spectrum can be very non-MFV like, while all the 
low-energy constraints will appear MFV-like. 

${\mathcal Z}_3^\chi$ is exactly conserved in many models of flavor. 
For instance, it remains unbroken for MFV new physics. 
More generally, ${\mathcal Z}_3^\chi$ remains exact if the flavor group 
${\cal G}_F^{\rm SM}$ is broken only by the vevs of scalar fields, or condensates, with zero flavor triality.
Examples of zero flavor triality fields are scalars in bifundamental or in adjoint 
representations of the flavor $SU(3)$'s.
The basic requirement for this set-up is that ${\cal G}_F^{\rm SM}$ is a good symmetry in the UV. 
This is achieved, if ${\cal G}_F^{\rm SM}$ is fully gauged in the UV, which is the possibility we explored. 
The DM is then stable because it is ${\mathcal Z}_3^\chi$ odd, while all SM fields are 
${\mathcal Z}_3^\chi$ even. 

We investigated two different types of flavored DM models: 
(i) models in which the leading interaction of the DM with the visible sector is through 
the flavored gauge bosons (FGBs), and (ii) models in which the contributions from the FGB 
exchanges are subleading.

As an example of the first type of models we considered a Dirac fermion DM that is in a 
fundamental representation of $SU(3)_U$. 
The relic abundance is fixed by the resonant DM annihilation to SM particles through the 
$s$-channel exchange of the lightest FGB. 
The DM is thus required to have a mass of about half of the lightest FGB's mass. 
This in turn implies that the FGBs cannot be arbitrarily heavy, but at most ${\mathcal O}(10$\,TeV). 
Such light FGBs are possible, if the masses of FGBs are inversely proportional to the corresponding 
quark masses. 
That is, if the FGBs that couple most strongly to light quarks are also the heaviest. 
To achieve this we adopted the model of Ref.~\cite{Grinstein:2010ve} in which the inverse proportionality
is achieved by introducing a set of quark partners also necessary to cancel gauge anomalies. 
The same quark partners also mix with the SM quarks and lead to the mass hierarchy of 
the SM quark masses.

The flavor and collider phenomenology of the model is very similar to the case where 
DM is not considered.
The fact that the first-generation quark partners are the heaviest and that 
the spectrum is completely split, signals the non-MFV character of the model.
However, the low-energy consequences are MFV-like (see Appendix~\ref{app:sec:mfv}).
The flavor constraints are satisfied even with FGBs and the top-quark partner, 
with masses potentially well below the TeV scale. 
The relevant direct collider searches are the searches for dijet resonances and $t'$ searches.
They exclude part of the available parameter space. 
Requiring that there is a thermal relic DM introduces new constraints. 
Because DM is part of a flavor multiplet the heavier DM states need to decay before big bang nucleosynthesis. 
In the case of radiatively-split fermionic DM this excludes a large part of the parameter space. 
The remaining points are mostly safe from direct-detection bounds. 
The fact that the DM mass is related to the FGB mass by the requirement of almost resonant
annihilation sets both lower and upper bounds on the DM mass. 
Requiring that the theory is perturbative also puts an upper bound on the DM mass, $m_{\chi_1}\lesssim 5$\,TeV. 
On the other hand, requiring that the FGBs satisfy flavor and direct constraints and that DM is simultaneously 
in accordance with cosmological constraints leads to a lower bound on the DM  
mass, $m_{\chi_1}\gtrsim 500$\,GeV.   
Improved bounds on dijet resonances at the LHC are expected to further strengthen 
this constraint (see Fig.~\ref{fig:dijet}). 

We have also considered the possibility that the DM multiplet is split due to an extra source of flavor breaking. 
Also in this case, the correct relic abundance requires resonant annihilation. 
The DM mass is in thus still roughly equal to half of the mass of the lightest FGB. 
However, the heavier DM states decay well before big bang nucleosynthesis so that a much wider 
range of DM masses is phenomenologically viable. 
In our scan this includes DM masses as light as $100$\,GeV (with very small couplings to FGBs) 
and up to $10$\,TeV.

A possible signal of the gauged flavor model with fermionic DM at the LHC are mono-jets, 
where the lightest FGB is produced associated with initial state radiation and decays to $\chi_1$ pairs. 
The $\chi_1$s are expected to be non-relativistic in the lightest FGB's rest frame, as 
their combined mass needs to be close to the FGB mass to fulfill the resonance condition 
for relic DM abundance. 
In the event that such a signal would eventually emerge, the corresponding dijet-resonance signal
is generically also expected in the model.
A final possibility in the case of radiatively split DM mass spectrum is that some of the 
lightest FGBs decay to slow-moving $\chi_{2,3}$. They in turn decay within the detector, 
leaving (highly) displaced vertices, isolated hits in the calorimeter or in the muon chambers. 
Unfortunately, in most of the parameter space $\chi_{2,3}$ are expected to decay well outside 
the detectors, see Fig.~\ref{fig:mDM-fermion3}, leaving mono-jets as the only signal.

In the second type of models, where FGB exchanges give only subleading contributions, 
the only visible consequence of the flavor dynamics on the DM is that DM is stable. 
The DM mass and the mass of the lightest FGB are no longer connected. 
We show this in the example of scalar flavored DM, in which the dominant interactions 
with the visible sector are through the Higgs portal operator. 
In this case the phenomenology of the DM is to a very good approximation the same as in 
the Higgs-portal scalar DM, while the dynamics of FGBs and quark partners is unrelated to DM.

In short, we have shown, using an explicit renormalizable model, that it is possible 
for flavored DM to be a thermal relic.
The considered model is not the only choice. 
One could consider DM in other representations of ${\cal G}_F^{\rm SM}$. 
Our analysis can be extended also in other ways: for instance, by enlarging the global symmetry
as in Ref.~\cite{Agrawal:2014aoa} and subsequently gauging it.  
For instance, with our field content the global group is ${\cal G}_F^{\rm SM}\times SU(3)_\chi$, 
where  $\chi$ is in the fundamental of $SU(3)_\chi$. 
In our work we have identified $SU(3)_\chi$ with $SU(3)_U$, but other choices could be made. 
Yet another possibility is to gauge only part of ${\cal G}_F^{\rm SM}$, for instance a
$U(2)^3 \subset {\cal G}_F^{\rm SM}$. 
Note that for fermionic DM, ${\mathcal Z}_3$ is part of an accidental $U(1)_\chi$ acting 
in the dark sector. 
The $U(1)_\chi$ can be broken by the dimension-$7$ operator $L H\chi\chi\chi$, 
but is exact in our renormalizable model. 
It can in principle be gauged and 
lead to additional phenomenology. If DM is a scalar, $U(1)_\chi$ can be broken already at the 
renormalizable level, leaving only $\mathcal Z_3$ exact.

{\bf Acknowledgements.} We thank Csaba Csaki, Gino Isidori, Graham Kribs and 
Christopher Smith  for enlightening discussions. 
J.Z.\, and F.B.\, are supported by the U.S. National Science Foundation under 
CAREER Grant PHY-1151392. This work was supported in part by the Slovenian Research Agency. F.B. is grateful for the hospitality of the Fermilab theory department. Fermilab is operated by Fermi Research Alliance, LLC under Contract No. DE-AC02-07CH11359 with the United States Department of Energy. The work of DR is supported by the NSF under grant No. PHY-1002399. 

\appendix
\section{Minimal flavor violation with gauged flavor symmetries}
\label{app:sec:mfv}

In this appendix we verify numerically that the Wilson coefficients in the 
weak Hamiltonian for $B_d$ and $B_s$ mixing, 
\beq
{\cal H}_{\rm eff}^{\Delta B=2}=\sum_{i=1}^5 C_i^{bq} Q_i^{bq}+\sum_{i=1}^5 \tilde C_i^{bq} \tilde Q_i^{bq},
\eeq
generated from exchanges of flavored gauge bosons, are of the MFV type. 
A tree-level exchange of FGBs generates contributions to $B_d$ mixing through operators 
\beq
Q_1^{bd}=\bar d_L^\alpha \gamma_\mu b^\alpha_L \bar d_L^\beta \gamma_\mu b^\beta_L, \qquad Q_3^{bd}=\bar d_R^\alpha  b^\beta_L \bar d_R^\beta b^\alpha_L,
\eeq
and $\tilde Q_1^{bd}, \tilde Q_3^{bd}$, that follow from $ Q_1^{bd}, Q_3^{bd}$ with 
$P_L\leftrightarrow P_R$ exchange (the remaining operators can be found in, e.g., Ref.~\cite{Bona:2007vi}). 
If the lightest FGB has predominantly left-handed couplings, then the $C_1^{bd}$ Wilson coefficient 
is the largest one. 
If the lightest FGB couples predominantly to the right-handed quarks, then $\tilde C_1^{bd}$ dominates. 
For comparable left- and right-handed couplings all four Wilson coefficients, 
$C_{1,3}^{bd}$, $\tilde C_{1,3}^{bd}$, are important. 
The analogous discussion applies to the case of FGB contributions to $B_s$ mixing 
obtained with a trivial $d\to s$ exchange.  

As discussed in Section~\ref{sec:model} the contributions from the gauged flavor model is
expandable in terms of the SM Yukawas. 
The contributions due to FGB exchanges can thus be written as
\beq \label{eq:C1bd:expand}
\delta C_1^{bd}=c_1  (y_u y_u^\dagger)_{13}^2 +c_2 (y_uy_u^\dagger)_{13} (y_d y_d^\dagger y_uy_u^\dagger )_{13}+\cdots=c_1 (V_{td}^*V_{tb})^2+c_2 y_d^2 (V_{td}^*V_{tb})^2+\cdots,
\eeq
where $(y_d)_{ij}=\diag(y_d,y_s,y_b)$, and we set $y_t=1$ in the second equality. 
In Eq.~\eqref{eq:C1bd:expand} we kept only the two terms relevant for the discussion below. 
The same expansion applies for $\delta C_1^{bs}$ with the replacement $d\to s$ in Eq.~\eqref{eq:C1bd:expand}.

\begin{figure}
  \begin{center}
    \includegraphics[height=.3\textwidth]{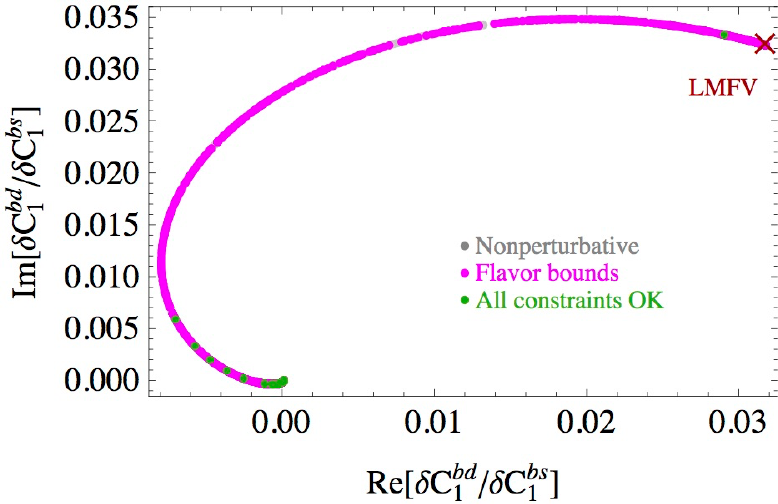}
    ~~~~~\includegraphics[height=.3\textwidth]{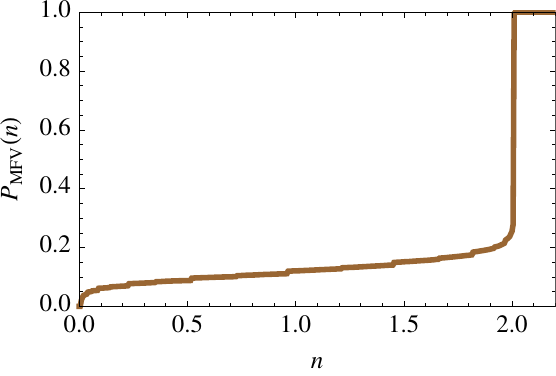}
    \\[-18mm]${}$
  \end{center}
  \caption{\label{fig:LLBsBd} The FGB contributions to the $V-A$ current operator 
  in the effective weak Hamiltonian. 
  The left panel shows the values of the complex ratio $\delta C_1^{bd}/\delta C_1^{bs}$ 
  for our scan points, with green points satisfying all constraints, magenta points excluded by 
  flavor constraints and grey points by perturbativity considerations. 
  The cross denotes the point $\delta C_1^{bd}/\delta C_1^{bs}=(V_{td}^*/V_{ts}^*)^2$.
  In this point the MFV operator with the smallest number of Yukawa insertions completely 
  dominates. 
  The right panel shows the cumulative function $P_{\rm MFV}(n)$, see Eq.~\eqref{eq:app:PMFV}.
  }
\end{figure}

In Fig.~\ref{fig:LLBsBd} (left) we show the ratio $\delta C_1^{bd}/\delta C_1^{bs}$, 
i.e. the NP contribution  to the $V-A$ quark current operator due to tree-level FGB exchanges. 
Note that the ratio $\delta C_1^{bd}/\delta C_1^{bs}$ can be complex. 
If $c_1$, i.e.\ the leading MFV (LMFV) term, dominates then 
$\delta C_1^{bd}/\delta C_1^{bs}\simeq (V_{td}^*/V_{ts}^*)^2$. 
This is denoted by a cross in Fig.~\ref{fig:LLBsBd} (left) . 
The addition of the operators with extra insertions of $y_d y_d^\dagger$ leads to  
$\delta C_1^{bd}/\delta C_1^{bs}$ not being equal to  $(V_{td}^*/V_{ts}^*)^2$. 
We verified that the curve for $\delta C_1^{bd}/\delta C_1^{bs}$ shown in Fig.~\ref{fig:LLBsBd} 
(left) can be fitted with the form of $\delta C_1^{bd,bs}$ in Eq.~\eqref{eq:C1bd:expand} by
taking $c_1$ real, $c_2$ complex, and varying $c_1$ from ${\mathcal O}(1)$ to vanishingly small. 
The points in our scan can be grouped into two sets. 
For the first set of points both $c_1$ and $c_2$ terms are sizeable. 
For the second set of points the $c_1$ term is negligible and $c_2$ dominates. 
This is shown in Fig.~\ref{fig:LLBsBd} (right), where we plot the cumulative 
distribution function
\beq
P_{\rm MFV}(n)=\frac{N\big(|\delta C_1^{bd}/\delta C_1^{bs}|\geq (m_d/m_s)^n |V_{td}^*/V_{ts}^*|^2\big)}{N_{\rm total}}.
\label{eq:app:PMFV}
\eeq
The function $P_{\rm MFV}(n)$ can be interpreted as the fraction of points that have 
the ratio $|\delta C_1^{bd}/\delta C_1^{bs}|$ effectively dominated by operators with 
up to $y_d^n$ insertions. 
That is, the points dominated by the $c_1$ term contribute to $P_{\rm MFV}(0)$ 
(and to $P_{\rm MFV}(n)$ with $n\geq0$), while the points dominated by the $c_2$ term 
contribute to $P_{\rm MFV}(2)$ (and to  $P_{\rm MFV}(n)$ with $n\geq 2$). 
The points with both $c_1$ and $c_2$ start contributing to $P_{\rm MFV}(n)$ for $n$ 
somewhere between $0$ and $2$, depending on the relative sizes of $c_1$ and $c_2$. 
Fig.~\ref{fig:LLBsBd} (right) shows that the $c_1$ term dominates in a subleading (but nonzero) 
set of points, that about 10\% points have both sizeable $c_1$ and $c_2$ terms, and that 
$c_2$ dominates in about $80\%$ of the points.

\begin{figure}
  \begin{center}
    \includegraphics[height=.29\textwidth]{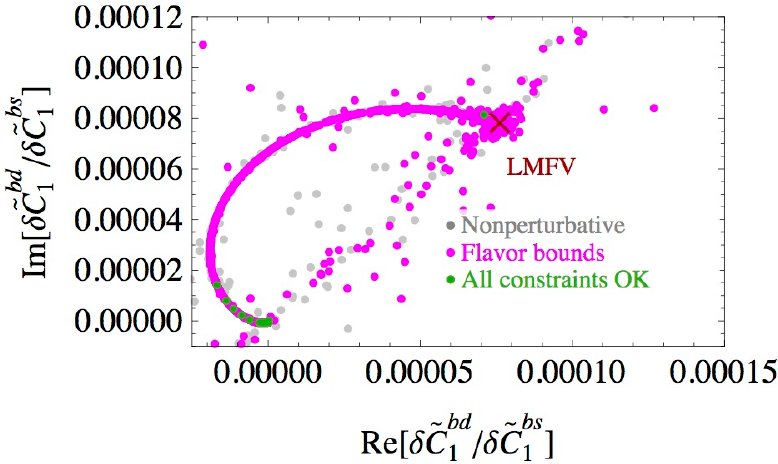}
   ~~~~~\includegraphics[height=.29\textwidth]{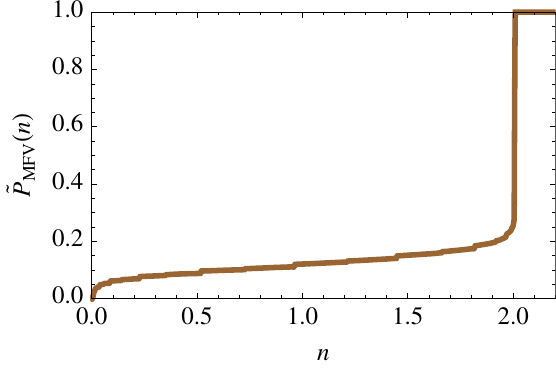}
    \\[-18mm]${}$
  \end{center}
  \caption{\label{fig:RRBsBd} 
 The FGB contributions to $V+A$ current operator in the effective weak Hamiltonian. 
 Left panel shows the complex ratio $\delta \tilde C_1^{bd}/\delta \tilde C_1^{bs}$ 
 for our scan points with the same color coding as in Fig.~\ref{fig:LLBsBd}. 
 The cross denotes the point $\delta \tilde C_1^{bd}/\delta \tilde C_1^{bs}=(m_d/m_s)^2 (V_{td}^*/V_{ts}^*)^2$, 
 obtained if the MFV operator with the smallest number of Yukawa insertions dominates. 
 The right panel shows the cumulative function $\tilde P_{\rm MFV}(n)$, 
 see Eq.~\eqref{eq:app:tildePMFV}.}
\end{figure}

The similar analysis can be performed for $V+A$ operators, $\tilde Q_1^{bd}$ and 
$\tilde Q_1^{bs}$. 
We expand the FGB contributions to their respective Wilson coefficients in terms of the SM Yukawas
\beq \label{eq:tildeC1bd:expand}
\begin{split}
\delta \tilde C_1^{bd}&=\tilde c_1  (y_d^\dagger y_u y_u^\dagger y_b)_{13}^2 +\tilde c_2 (y_d^\dagger y_uy_u^\dagger y_b )_{13} (y_d^\dagger y_d y_d^\dagger y_uy_u^\dagger y_b )_{13}+\cdots\\
&=\tilde c_1y_d^2 y_b^2 (V_{td}^*V_{tb})^2+\tilde c_2 y_d^4 y_b ^2(V_{td}^*V_{tb})^2+\cdots,
\end{split}
\eeq
and similarly for $\delta \tilde C_1^{bs}$ with the $d\to s$ replacement. 
We show in Fig.~\ref{fig:RRBsBd} (left) the ratio $\tilde C_1^{bd}/\tilde C_1^{bs}$ for our scan.  
If the $\tilde c_1$ term dominates, then $\tilde C_1^{bd}/\tilde C_1^{bs}=(m_d/m_s)^2 (V_{td}^*/V_{ts}^*)^2$, 
which is denoted by the cross in  Fig.~\ref{fig:RRBsBd} (left). 
The points for which also the $\tilde c_2$ operator (and other operators denoted by ellipses above) 
is important then lie away from the  $\tilde C_1^{bd}/\tilde C_1^{bs}=(m_d/m_s)^2 (V_{td}^*/V_{ts}^*)^2$ region. 
We also define a cumulative function 
\beq
\tilde P_{\rm MFV}(n)=\frac{N\big(|\delta \tilde C_1^{bd}/\delta \tilde C_1^{bs}|\geq (m_d/m_s)^{n+2} |V_{td}^*/V_{ts}^*|^2\big)}{N_{\rm total}}.\label{eq:app:tildePMFV}
\eeq
The values for $\tilde P_{\rm MFV}(n)$ are shown in Fig.~\ref{fig:RRBsBd} (right). 
We see that also in this case the points cluster into two groups, with vanishing 
$\tilde c_1$ term or with both $\tilde c_1$ and $\tilde c_2$ relevant. 

The above analysis demonstrates that the FGB contributions to the Wilson coefficients 
in the effective weak Hamiltonian can be expanded in terms of the SM Yukawas.
This is a hallmark of (general) MFV. 
In particular, the expansion in terms of $m_{d,s}/m_b$ and off-diagonal CKM elements 
can still be performed and is not ruined by the large ratios of scales present in the 
problem such as the very disparate FGB masses.

%==============================================================================
\section{Thermal relic computation}
\label{app:sec:thermalrelic}
%==============================================================================
\newcommand{\avg}[1]{\left\langle #1 \right\rangle}
\newcommand{\vlab}{v_{\rm lab}}

In this appendix we describe the calculation of relic density that was used in the 
scans in the main part of the paper. 
Several approximations to the coupled Boltzmann equations were necessary in order 
to reduce the evaluation time per benchmark and thus allow adequate coverage of the 
parameter space. We find the approximate solutions to be in agreement with the full 
solutions at the $\mathcal{O}(30\%)$ level.
The full numerical solution of the Boltzmann equations was obtained with 
{\tt MadDM}~\cite{Backovic:2013dpa} using a \texttt{UFO} model file~\cite{Degrande:2011ua},
which was generated with the \texttt{FeynRules} package~\cite{Alloul:2013bka}.

Denoting the DM multiplet by $\varphi$, where $\varphi$ is either a Dirac fermion 
or a complex scalar, the most general set of coupled Boltzmann equations reads
\cite{Gondolo:1990dk}
\begin{equation}
\begin{split}
\frac{dn_{\varphi_i}}{dt}+3Hn_{\varphi_i}=&
	-\sum_j\avg{\sigma(\varphi_i\varphi_j\leftrightarrow XX)\,\vlab}\left(n_{\varphi_i}n_{\varphi_j}-n_{\varphi_i}^\text{\sc eq} n_{\varphi_j}^\text{\sc eq}\right)\\
	& -\sum_{j\neq i}\avg{\sigma(\varphi_iX\leftrightarrow \varphi_j X)\vlab}
		\left( n_{\varphi_i}-\frac{n_{\varphi_i}^\text{\sc eq}}{n_{\varphi_j}^\text{\sc eq}}n_{\varphi_j}\right) n_X^\text{\sc eq}\\
	& -\sum_{j\neq i}\avg{\sigma(\varphi_i\varphi_j\leftrightarrow \varphi_k \varphi_\ell)\vlab}
		\left( n_{\varphi_i}n_{\varphi_j}
		-\frac{n_{\varphi_i}^\text{\sc eq}n_{\varphi_j}^\text{\sc eq}}
			{n_{\varphi_k}^\text{\sc eq}\varphi_\ell^\text{\sc eq}}
			 n_{\varphi_k}n_{\varphi_\ell}\right)\\
	& \pm \sum_{j\neq i}\left[ \avg{\Gamma(\varphi_{j,i}\rightarrow \varphi_{i,j}X)}n_{\varphi_{j,i}}
		+ \avg{\sigma(\varphi_{j,i}X\rightarrow \varphi_{i,j})}n_{\varphi_{j,i}}n_X^\text{\sc eq}\right],
\end{split}
\label{eq:BEGN}
\end{equation}
where $X$ denotes a generic SM state.
For large mass splittings between the $\varphi$ components it is sufficient to consider 
the lightest $\varphi_i$ state in Eq.~\eqref{eq:BEGN}. 
The contributions to the DM relic density from the heavy $\varphi$ components  
are exponentially suppressed by corresponding Boltzmann factors and can be 
neglected within our precision.
In contrast, when the mass splittings are small the full set of coupled Boltzmann 
equations in Eq.~\eqref{eq:BEGN} needs to be considered. 
Nevertheless, even in this case several approximations are possible for our model, 
as we explain below.

\begin{figure}
\includegraphics[scale=1]{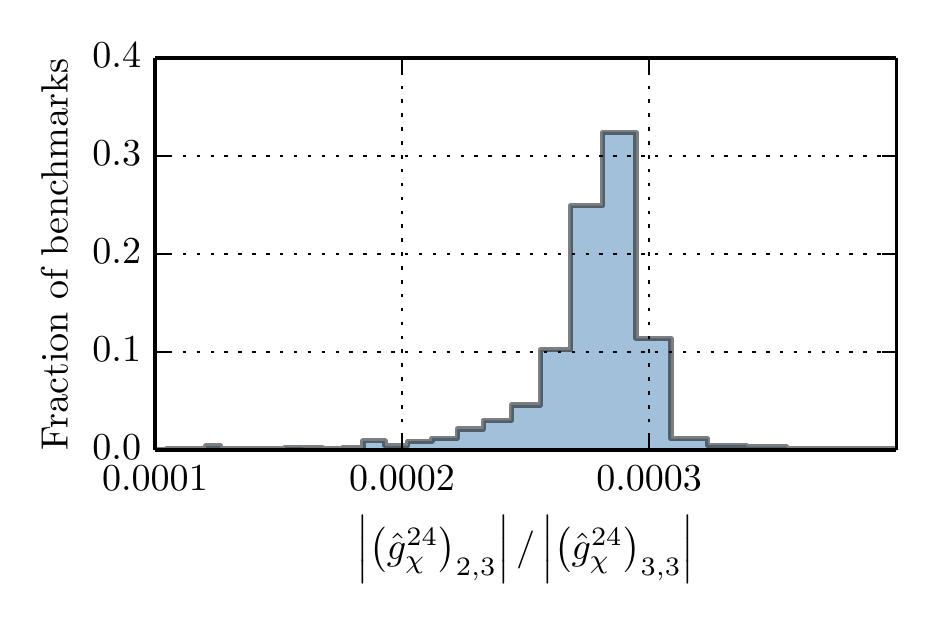}
\caption{
The fraction of benchmarks as a function of the off-diagonal couplings of the heaviest and next-to-heaviest 
DM components 
to the lightest FGB ($A^{24}$) normalized by the diagonal coupling of the heaviest component. 
}
\label{fig:ODHIST}
\end{figure}

First of all, the coannihilation of different $\varphi_i$ components into SM particles, 
$\varphi_i \varphi_j \to XX$ ($i\ne j$), can be safely neglected in our model. 
In benchmarks that survive the experimental constraints the off-diagonal couplings 
of the lightest FGB to $\varphi$ are much smaller than the diagonal ones, see 
Fig.~\ref{fig:ODHIST}. 
Secondly, in the calculation of DM relic density we also neglect the flavor-changing 
DM scattering off the thermal background, $\varphi_i X\to \varphi_j X$.
The $\varphi_i X\to \varphi_j X$ scattering can be important if 
$\avg{\sigma(\varphi_iX\rightarrow \varphi_j X)\vlab} / \avg{\sigma(\varphi_i\varphi_j\rightarrow XX)\,\vlab} \gtrsim n_{\varphi_j}/n_X^\text{\sc eq} \sim 10^{-9}$. 
In this case the off-diagonal couplings of $\mathcal{O}(10^{-4})$ relative to the 
diagonal ones are in principle large enough to have $\mathcal{O}(1)$ effects on the relic density, 
and neglecting $\varphi_i X\to \varphi_j X$ may not be justified. 
Therefore, for the benchmarks with  $(\hat g_\chi^{24})_{23}/(\hat g_\chi^{24})_{33}>3 \times 10^{-4}$ and small mass 
splittings among $\varphi$, we explicitly verified using \texttt{MadDM} 
that neglecting $\varphi_i X\to \varphi_j X$ scattering 
leads to a change in DM relic density smaller than ${\mathcal O}(30\%)$.

Finally, we are able to neglect the pure DM scattering process in the third line 
of Eq.~\eqref{eq:BEGN} since $\avg{\sigma(\varphi_i\varphi_j\leftrightarrow \varphi_k \varphi_\ell)\vlab}\ll \avg{\sigma(\varphi_i\varphi_j\leftrightarrow XX)\,\vlab}$ in our model. 
The largest contribution to this process is from diagonal couplings between the FGB and DM. 
This process can couple the evolution of the DM species if 
$\avg{\sigma(\varphi_i\varphi_j\leftrightarrow \varphi_k \varphi_\ell)\vlab}\sim\avg{\sigma(\varphi_i\varphi_j\leftrightarrow XX)\,\vlab}$. 
The diagonal FGB couplings to the  quarks and to the DM of the same generation 
are approximately equal.
By accounting for the color factors and the multiplicity of channels when annihilating 
into SM fields one concludes that the pure DM scattering is indeed subleading.

Therefore, for almost mass degenerate $\varphi_i$ it is sufficient to consider a set of uncoupled Boltzmann equations
\beq\label{eq:relicDM:decoupled}
\frac{dn_{\varphi_i}}{dt}+3Hn_{\varphi_i}=
	-\sum_j\avg{\sigma(\varphi_i\varphi_i\leftrightarrow XX)\,\vlab}\left(n_{\varphi_i}^2 -n_{\varphi_i}^\text{\sc eq}{}^2 \right).
\eeq
The DM relic abundance is in this case the sum of relic abundances for each 
of the three components obtained from the above set of equations 
(the heavy $\varphi$ components, in our case $\varphi_{2}$ and $\varphi_3$,  
decay after their respective freeze-outs and contribute to the $\varphi_1$ DM relic abundance).
In contrast, for large mass splittings the heavy $\varphi$ components are irrelevant 
for the calculation of the DM relic abundance. 
This is then obtained from Eq.~\eqref{eq:relicDM:decoupled} by considering only the 
lightest DM state, in our case $\varphi_1$.

We calculate the DM relic abundance from Eq.~\eqref{eq:relicDM:decoupled} using the 
freeze-out approximation~\cite{Griest:1990kh}, which gives
\begin{equation}
\Omega h^{2}=\frac{1.07\times10^{9}\,\textrm{GeV}^{-1}}{J(x_{f})\sqrt{g_{*}}M_{Pl}}\,,
\end{equation}
where $M_{Pl}=1.22\times10^{19}\,$\,GeV, $g_{*}$ is the total number
of effectively relativistic degrees of freedom at the time of the
freeze-out, and
\begin{equation}
J(x_{f})=\int_{x_{f}}^{\infty}dx\frac{\left\langle \sigma v_{\rm lab}\right\rangle _{\textrm{th}} }{x^{2}}\,.
\label{eq:j}
\end{equation}
The freeze-out temperature ($x_{f}=m_{\varphi_{1}}/T_{f}$)
is obtained by solving
\begin{equation}
x_{f}=\ln\frac{0.038\,g_{\textrm{eff}}\,M_{Pl}\,m_{\varphi_{1}}\left\langle \sigma v_{\rm lab}\right\rangle _{\textrm{th}} }{\sqrt{g_{*}x_{f}}}\,,
\end{equation}
where the thermally-averaged cross section is
\begin{equation}
\left\langle \sigma v_{\textrm{lab}}\right\rangle _{\textrm{th}}=\frac{2x^{\frac{3}{2}}}{\sqrt{\pi}}\int_{0}^{\infty} \sigma_{\rm eff} v_{\textrm{lab}}\,\sqrt{\epsilon}~e^{-x\epsilon}~d\epsilon\,,
\end{equation} 
with $v_{\textrm{lab}}=2\sqrt{\epsilon(1+\epsilon)}/(1+2\epsilon)$ and $\epsilon=s/(2m_{\varphi_{1}})^{2}-1$. 
The freeze-out approximation is accurate to a few percent with respect 
to the full numerical solution of the Boltzmann equation~\cite{Gondolo:1990dk}.

The fermionic flavored DM annihilates through the s-channel exchange of FGBs. 
In this case, the integration over $x$ can be performed analytically and the 
double integral in Eq.~\eqref{eq:j} reduces to a single one that can be efficiently 
evaluated numerically. 
In particular,
\begin{equation}
J(x_{f})=\int_{0}^{\infty} 2 \sigma v_{\rm lab} \textrm{Erfc}(\sqrt{x_f \epsilon})~d\epsilon~.
\end{equation}
We evaluate the above integral numerically in the parameter scan.

%==============================================================================

In the annihilation cross section of the fermionic flavored DM we keep the 
dominant contribution -- the $s$-channel exchange of the lightest FGB, $A^{24}$. 
The annihilation cross section for $\chi_i\bar{\chi}_i\to\bar{u}_j u_j$ 
(and similarly for $\chi_i\bar{\chi}_i\to\bar{d}_j d_j$ ) is given by
\beq\label{app:eq:annihilation:fDM}
\begin{split}
\sigma(\chi_i\bar{\chi}_i\to\bar{u}_j u_j)=&\frac{(\hat g_{\chi}^{24})_{ii}^2 }{4\pi}\sqrt{\frac{s-4m_{u_j}^{2}}{s-4m_{\chi_i}^{2}}}\left(1+\frac{2m_{\chi_i}^{2}}{s}\right)\times\\
&\times\frac{\big(\hat {\cal G}_V^{u}\big)_{jj,24}^2 \left(s+2m_{u_j}^{2}\right)+\big(\hat {\cal G}_A^{u}\big)_{jj,24}^2\left(s-4m_{u_j}^{2}\right)}{\left(s-m_{A^{24}}^{2}\right)^{2}+m_{A^{24}}^{2}\Gamma_{A^{24}}^{2}},
\end{split}
\eeq
where the vector and axial-vector couplings to quarks, $\hat{\cal G}_{A,V}$, were 
defined in Eq.~\eqref{eq:GVA:def},
$\sqrt{s}$ is the center-of-mass energy and  $\Gamma_{A^{24}}$ is the total 
decay width of the lightest FGB. 
The decay rate for $A^{24}\to \bar{u}_j u_j$ assuming $m_{A^{24}}>2m_{u_j}$ is
\beq\label{app:eq:A24:decay:width}
\Gamma(A^{24}\to\bar{u}_j u_j)=\frac{m_A^{24}}{4\pi}\sqrt{1-\frac{4m_{u_j}^{2}}{m_{A^{24}}^{2}}}
\left[\big(\hat {\cal G}_V^{u}\big)_{jj,24}^2 \left(1+\frac{2m_{u_j}^{2}}{m_{A^{24}}^{2}}\right)+\big(\hat {\cal G}_A^{u}\big)_{jj,24}^2\left(1-\frac{4m_{u_j}^{2}}{m_{A^{24}}^{2}}\right)\right].
\eeq
The rate for $A^{24}\to\chi_i \bar{\chi}_i,\bar{d}_j d_j$ is obtained after trivial 
replacements for masses and couplings (and dividing by the $N_c$ color factor 
for decays to $\chi_i \bar{\chi}_i$). 
The total FGB decay rate is obtained after summing over all kinematically allowed decay channels.

%==============================================================================
\section{Higgs coupling Feynman rules}
\label{app:feynrules}
%==============================================================================
As noted in Sec.~\ref{sec:direct}, the $h\bar{f}f$ Feynman rules given in Appendix 
A.1 of Ref.~\cite{Buras:2011wi} contain a typo. 
The corrected Feynman rules are given here.
\begin{equation}
\parbox[c]{4cm}{\includegraphics[width=4cm]{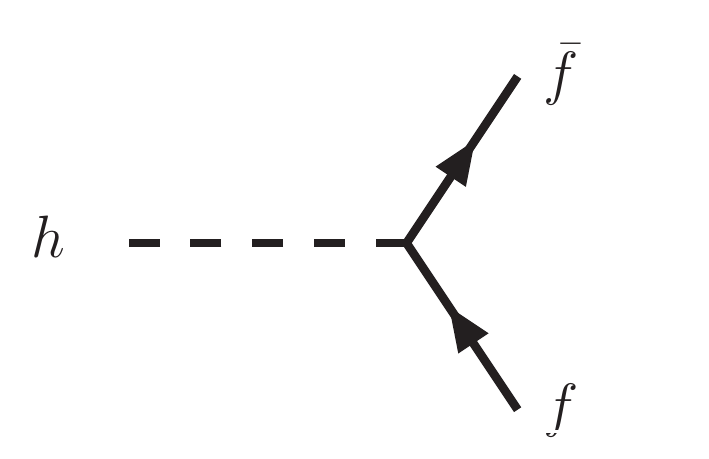}}= \frac{i}{\sqrt{2}}\left( C_L\,P_L+C_R\,P_R \right)
\label{eq:hff-fr}
\end{equation}
where the couplings $C_L$ and $C_R$ are:
\begin{equation}
\begin{split}
h\bar{u}_i u_i\quad&: C_L=C_R=+\lambda_u\,s_{u_Ri}\,c_{u_Li}\\
h\bar{u}_i^\prime u_i^\prime\quad&: C_L=C_R=-\lambda_u\,c_{u_Ri}\,s_{u_Li}\\
h\bar{u}_i u_i^\prime \quad&: \Bigg\{
  \begin{matrix}
    C_R = -\lambda_u\, c_{u_Ri}\,c_{u_Li}\\
    C_L = +\lambda_u\, s_{u_Ri}\,s_{u_Li}
  \end{matrix}\\
h\bar{u}_i^\prime u_i \quad&: \Bigg\{
  \begin{matrix}
    C_R = +\lambda_u\, s_{u_Ri}\,s_{u_Li}\\
    C_L = -\lambda_u\, c_{u_Ri}\,c_{u_Li}
  \end{matrix}
\label{eq:hff-couplings}
\end{split}
\end{equation}

\end{document}